\documentclass[aps,prl,reprint,superscriptaddress, noeprint]{revtex4-2}
\usepackage[utf8]{inputenc}
\usepackage{natbib}

\usepackage{amsmath}
\usepackage[toc,page]{appendix}
\usepackage{bm}
\usepackage{ amssymb }
\usepackage{graphicx}
\usepackage {epstopdf}
\usepackage{csquotes}
\usepackage{float}
 \usepackage{color, soul}

\begin{document}

\title{Fabry-P{\'e}rot interferometry at the $\nu$ = 2/5 fractional quantum Hall state}

\author{J. Nakamura}
\affiliation{Department of Physics and Astronomy, Purdue University, West Lafayette, Indiana, 47907, USA}
\affiliation{Birck Nanotechnology Center, Purdue University, West Lafayette, Indiana, 47907, USA}


\author{S. Liang}
\affiliation{Department of Physics and Astronomy, Purdue University, West Lafayette, Indiana, 47907, USA}
\affiliation{Birck Nanotechnology Center, Purdue University, West Lafayette, Indiana, 47907, USA}

\author{G. C. Gardner}
\affiliation{Birck Nanotechnology Center, Purdue University, West Lafayette, Indiana, 47907, USA}
\affiliation{Microsoft Quantum Lab West Lafayette, West Lafayette, Indiana, 47907, USA}

\author{M. J. Manfra}
\email[]{mmanfra@purdue.edu}
\affiliation{Department of Physics and Astronomy, Purdue University, West Lafayette, Indiana, 47907, USA}
\affiliation{Birck Nanotechnology Center, Purdue University, West Lafayette, Indiana, 47907, USA}
\affiliation{Microsoft Quantum Lab West Lafayette, West Lafayette, Indiana, 47907, USA}
\affiliation{Elmore Family School of Electrical and Computer Engineering, Purdue University, West Lafayette, Indiana, 47907, USA}
\affiliation{School of Materials Engineering, Purdue University, West Lafayette, Indiana, 47907, USA}

\date{\today}

\begin{abstract}
Electronic Fabry-P{\'e}rot interferometry is a powerful method to probe quasiparticle charge and anyonic braiding statistics in the fractional quantum Hall regime. We extend this technique to the hierarchy $\nu = 2/5$ fractional quantum Hall state, possessing two edge modes that in our device can be interfered independently. The outer edge mode exhibits interference similar to the behavior observed at the $\nu = 1/3$ state, indicating that the outer edge mode at $\nu = 2/5$ has properties similar to the single mode at $\nu = 1/3$. The inner mode shows an oscillation pattern with a series of discrete phase jumps indicative of distinct anyonic braiding statistics. After taking into account the impact of bulk-edge coupling, we extract an interfering quasiparticle charge ${e^*} = 0.17 \pm 0.02$ and anyonic braiding phase $\theta _a = (-0.43 \pm 0.05)\times 2\pi$, which serve as experimental verification of the theoretically predicted values of $e^* = \frac{1}{5}$ and $\theta _a = -\frac{4\pi}{5}$.  

\end{abstract}
\date{\today}

\maketitle

\section{Introduction}

The fractional quantum Hall effect \cite{Tsui1982} is the archetype of a strongly interacting topological phase of matter hosting anyonic excitations. Quantum Hall states at fractional filling factors of the form $\nu = \frac{1}{2p+1}$ are described by the Laughlin wave function \cite{Laughlin1983}. Higher order fractional states $\nu = \frac{n}{2pn+1}$ ($n,p$ integers) may be understood in terms of the hierarchical construction \cite{Haldane1983, Halperin1984} or the composite fermion model \cite{Jain1989, Jain1990, JainBook}. The elementary excitations of fractional quantum Hall states are quasiparticles carrying a fraction of an electron's charge \cite{Laughlin1983} and obeying anyonic braiding statistics \cite{Leinaas1977, Goldin1980, Wilczek1982, Halperin1984, Arovas1984}. At the $\nu = 1/3$ Laughlin state, fractional charge has been observed through resonant tunneling \cite{Goldman1995, Roosli2021}, shot noise \cite {Saminadayar1997, Reznikov1997}, scanning SET techniques \cite{Martin2004}, and interference \cite{Nakamura2019}, while evidence for anyonic statistics has been observed with Fabry-P{\'e}rot interferometry \cite{Nakamura2020, Nakamura2022} and in quasiparticle collision experiments \cite{Bartolomei2020, Ruelle2023, Glidic2023}. 

It is natural to attempt to extend experimental probes of exotic statistics to the hierarchy state $\nu = 2/5$, one of the principal daughter states of $\nu=1/3$. Recent shot noise cross-correlation experiments have reported novel behavior at $\nu = 2/5$ \cite{Glidic2023, Ruelle2023}, providing evidence for sensitivity to statistical properties of $e/5$ quasiparticles, where $e$ is the charge of an electron. Two-particle time domain shot-noise experiments indicate that quantum coherence can be maintained for $e/5$ anyons at $\nu = 2/5$ \cite{Taktak2022}. 
Electronic Fabry-P{\'e}rot interferometry is a powerful probe of quasiparticle charge and statistics, and has been studied in numerous theoretical \cite{Kivelson1989, Kivelson1990, Chamon1997, Kim2006, Bonderson2006, Stern2006, Stern2008, Rosenow2007, Stern2010, VonKeyserlingk2015, Feldman2021, Carrega2021, Feldman2022} and experimental works \cite{Ji2003, Deviatov2008, Roulleau2008, Zhang2009, McClure2009,Ofek2010, McClure2012a, Kou2012, Willett2009, Willett2013, Willett2023}, with recent experiments extending the measurement technique to the quantum Hall effect in graphene \cite{Ronen2021, Deprez2021, Zhao2022}.

Here we describe the operation of a Fabry-P{\'e}rot interferometer and quantitative analysis of braiding statistics at $\nu=2/5$. There are challenges to extending interferometery to more fragile, higher-order states. The energy gap at $\nu=2/5$ is significantly smaller than at $\nu=1/3$, making it important to use a high-quality heterostructure with reasonably high electron density so that an incompressible state can be achieved both in the bulk 2DEG and in a confined device. Also, $\nu = 2/5$ is expected to have two distinct charged edge modes, making it important to independently interfere {\it both} the inner and outer modes. At $\nu=2/5$, the bulk quasiparticles and the tunneling charge on the inner edge mode are expected to carry fractional charge $e/5$ \cite{Jain1990, JainBook, Halperin2011, Reznikov1999}. The expected value of the anyonic phase for braiding one $e/5$ quasiparticle around another (or exchanging positions twice) is $\theta _a = -2\pi \times \frac{2p}{2pn+1}$, yielding $\theta _{a} = -2\pi \times \frac{2}{5}$ \cite{Halperin1984, Su1986, Jeon2003a, Halperin2011}.

\begin{figure}[h]
\def\ffile{SEM_nu_1_v3.pdf}
\centering
\includegraphics[width=1.0\linewidth]{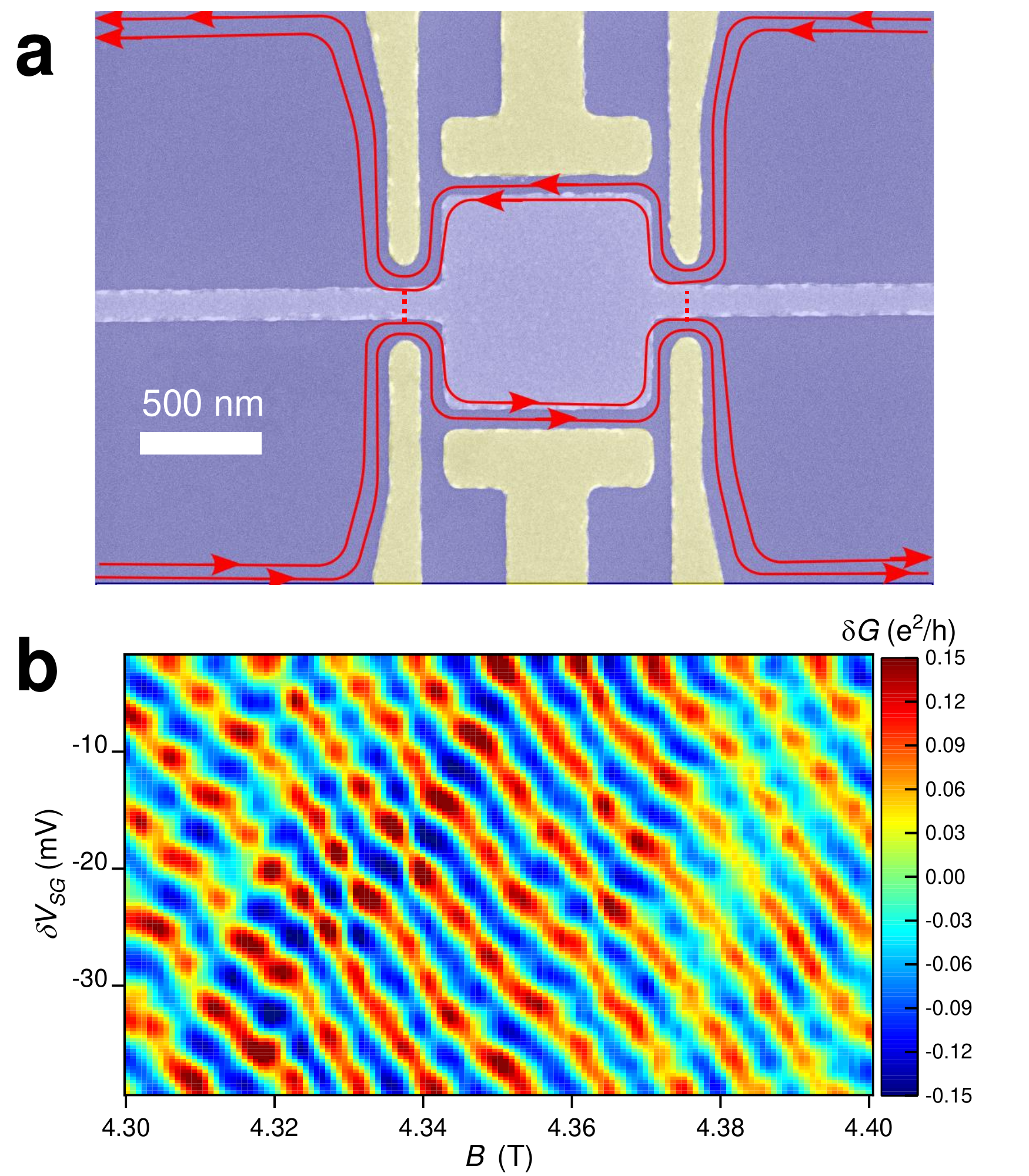}
\caption{\label{Device}a) False-color SEM image an interferometer with the same dimensions as the device used in our experiment. Red lines indicate propagating edge states. b) Conductance oscillations at $\nu=1$ versus magnetic field and side gate voltage $V_{SG}$. The negatively sloped lines of constant phase indicate Aharonov-Bohm regime behavior, while weak modulations suggest small but finite bulk-edge coupling.}
\end{figure}

A quantum  Hall Fabry-P{\'e}rot interferometer uses two quantum point contacts (QPCs) to partially reflect incident edge states which propagate around the gate-defined edge of the 2DEG. Interference occurs between coherent waves backscattered at each of the QPCs. In the limit of weak backscattering, the quantum Hall interferometer generates a phase difference given by Eqn. \ref{theta_AB} \cite{Chamon1997, Halperin2011}:
\begin{equation} \label{theta_AB}
\frac{\theta}{2\pi}=e^* \frac{AB}{\Phi_0}+N_{qp} \frac{\theta_a}{2\pi}
\end{equation}
Note that this interference phase is for counter-clockwise propagation of a quasiparticle around the loop with magnetic field in the $-\hat{z}$ direction. Eqn. \ref{theta_AB} includes the Aharonov-Bohm (AB) phase and the anyonic phase contribution $\theta_a$, with $N_{qp}$ the number of quasiparticles localized inside the interferometer. $e^*$ is the dimensionless ratio of the quasiparticle charge to the bare electron charge, $A$ is the area of the interference path around which the edge states circulate; this area is defined by the QPC gates and side gates. For $\nu=1/3$, theory predicts $e^*=\frac{1}{3}$ and $\theta_a= \frac{2\pi}{3}$, while for $\nu = 2/5$, $e^* = \frac{1}{5}$ and $\theta _a=\frac{-4\pi}{5}$ \cite{Arovas1984, Halperin1984, Blok1990, Jeon2003a, Jeon2004, Stern2008, Halperin2011}. Note that the charge $e/5$ at $\nu = 2/5$ is associated with localized quasiparticles in the bulk and the tunneling charge on the inner edge mode. The outer edge, belonging to the lowest composite fermion Lambda level \cite{Jain1989, JainBook}, is expected to support charge $e/3$ quasiparticles, as at $\nu = 1/3$. This has been confirmed in shot noise experiments \cite{Reznikov1999}.   

If the chemical potential is in an energy gap so that the bulk is incompressible and the quasiparticle number is fixed as $B$ and $V_{SG}$ are varied, then the conductance will oscillate with a flux period of $\frac{\Phi_0}{e^*}$. At fixed filling factor, increasing magnetic field increases the number of electrons in the quantum Hall condensate $q_{condensate}= \frac{\nu A B}{\Phi_0}$. At specific values of $B$ and $V_{SG}$, local variations in the disorder potential landscape may favor the addition of a hole-like quasiparticle (or the removal of an electron-like quasiparticle) inside the bulk rather than the continuous addition of charge that keeps the filling factor fixed, leading to a discrete change in phase by $-\theta _a$. Note that $\nu = 2/5$ has smaller quasiparticle charge compared to $\nu = 1/3$, making the Aharonov-Bohm phase evolve more gradually with changes in magnetic field, and a larger magnitude of $\theta _a$; these two factors will make the discrete jumps in phase when the quasiparticle number changes a more dramatic effect than at $\nu=1/3$, since the anyonic phase contribution is much larger relative to the slowly varying Aharonov-Bohm component. 

Eqn. \ref{theta_AB} neglects the effects of bulk-edge coupling, which can cause the area of the interference path to change when charge in the bulk changes (either in the condensate or in the form of localized quasiparticles). In the presence of finite bulk-edge coupling, the interference phase will be modified \cite{Rosenow2007, Halperin2011, VonKeyserlingk2015}:

\begin{equation} \label{theta_BE}
\frac{\theta}{2\pi}=e^* \frac{\Bar{A} B}{\Phi _0}- \kappa \frac{e^*}{\Delta \nu} (e^*N_{qp} + \nu_{in} \frac{\Bar{A} B}{\Phi _0}-\Bar{q_b})+N_{qp} \frac{\theta_a}{2\pi}
\end{equation}

In Eqn. \ref{theta_BE}, $\Bar{A}$ is the ideal area of the interference path (not including variations $\delta A$ due to the bulk-edge coupling), $\Delta \nu$ is the difference between the filling factor corresponding to the interfering edge state and the filling factor of the next-outer fully transmitted edge state, $\nu_{in}$ is the filling factor corresponding to the interfering edge state, and $\Bar{q_b}$ is the charge of ionized impurities and induced charge on the metallic gate resulting  from the applied gate voltages. $\kappa \equiv -\frac{\delta q_i}{\delta q_b}$ is the effective bulk-edge coupling parameter which describes how much the charge on the interfering edge mode ($q_i$) changes (and thus area) in response to a change in bulk charge (the total excess bulk charge is the term in parantheses in Eqn. \ref{theta_BE},  $\delta q_b\equiv e^*N_{qp} + \nu_{in} \frac{\Bar{A} B}{\Phi _0}-\Bar{q_b}$). Note that $q_b$ and $q_i$ are measured in units of the bare electron charge $e$.

\section{Measurements at $\nu=1$}
The primary device studied in this work, labeled Device A, consists of two quantum point contacts (QPCs) which partially reflect edge modes, and a pair of side gates to define an interference path. An SEM image of an identical device is shown in Fig. \ref{Device}a. This device has lithographic dimensions 1 $\mu$m $\times 1$ $\mu$m, and is fabricated on a GaAs/AlGaAs heterostructure utilizing the screening well design similar to our previous experiments \cite{Nakamura2019}; this heterostructure reduces bulk-edge coupling, making it possible to observe the anyonic phase when the localized quasiparticle number changes \cite{Nakamura2020}. The degree of residual bulk-edge coupling is controlled through heterostructure and device design \cite{Nakamura2022}. The heterostructure and device are detailed in Supp. Fig. 1 and Supp. Section 1 (simulations were done with the Nextnano software \cite{Birner:2007}). The electron density is $n\approx 1.05\times 10^{11}$ cm$^{-2}$ and mobility $\mu \approx 9 \times 10^6$ cm$^2$/Vs. Note that this density is higher than in previous devices used to probe braiding at $\nu = 1/3$ \cite{Nakamura2020, Nakamura2022}; higher density enables a more robust $\nu = 2/5$ state. Conductance measurements are made using standard lock-in amplifier techniques with a typical excitation voltage of 10$\mu$V and frequency of 13 Hz, and are performed in a dilution refrigerator at a temperature of $T$=10mK except where otherwise noted.

Fig. \ref{Device}b shows conductance variation $\delta G$ (with a smooth background subtracted) versus magnetic field $B$ and side gate voltage variation $\delta V_{SG}$ at the integer quantum Hall state $\nu = 1$. Note that $\Delta V_{SG}$ is relative to -1.8 V in all measurements, and the side gate voltage excursions around this value are small so that the area changes by only a few percent. Since at $\nu=1$ the dominant tunneling charge is the electron, the interference phase will be determined by the AB phase in Eqn. \ref{theta_AB}, and the oscillation period will be $\Phi_0$, making it possible to extract the effective area of the interferometer and the lever arm $\frac{\partial \Bar{A}}{\partial V_{SG}}$ relating a change in gate voltage to change in area (See Supp. Section 2 and Supp. Fig. 2). Negatively sloped lines of constant phase are a signature of Aharonov-Bohm interference \cite{Rosenow2007, Halperin2011, Zhang2009}, while weak modulations in the pattern suggest small but finite bulk-edge coupling. The magnetic field period $\Delta B = 11$ mT gives the effective area $A = \frac{\Phi_0}{\Delta B} \approx 0.38$ $\mu$m$^2$. This indicates an approximate depletion length of 200 nm around the gates, consistent with previous measurements and simulations of similar devices \cite{Nakamura2019, Nakamura2020, Nakamura2022}. The gate voltage oscillation period is $\Delta V_{SG} = 8.5$ mV, yielding $\frac{\partial \Bar{A}}{\partial V_{SG}} = \frac{\Phi_0}{B \Delta V_{SG}}=0.11$ $\mu$m$^2$V$^{-1}$.

\section{Measurements at $\nu = 2/5$}

\begin{figure}[h]
\def\ffile{RD_and_QPCs_v2.pdf}
\centering
\includegraphics[width=1.0\linewidth]{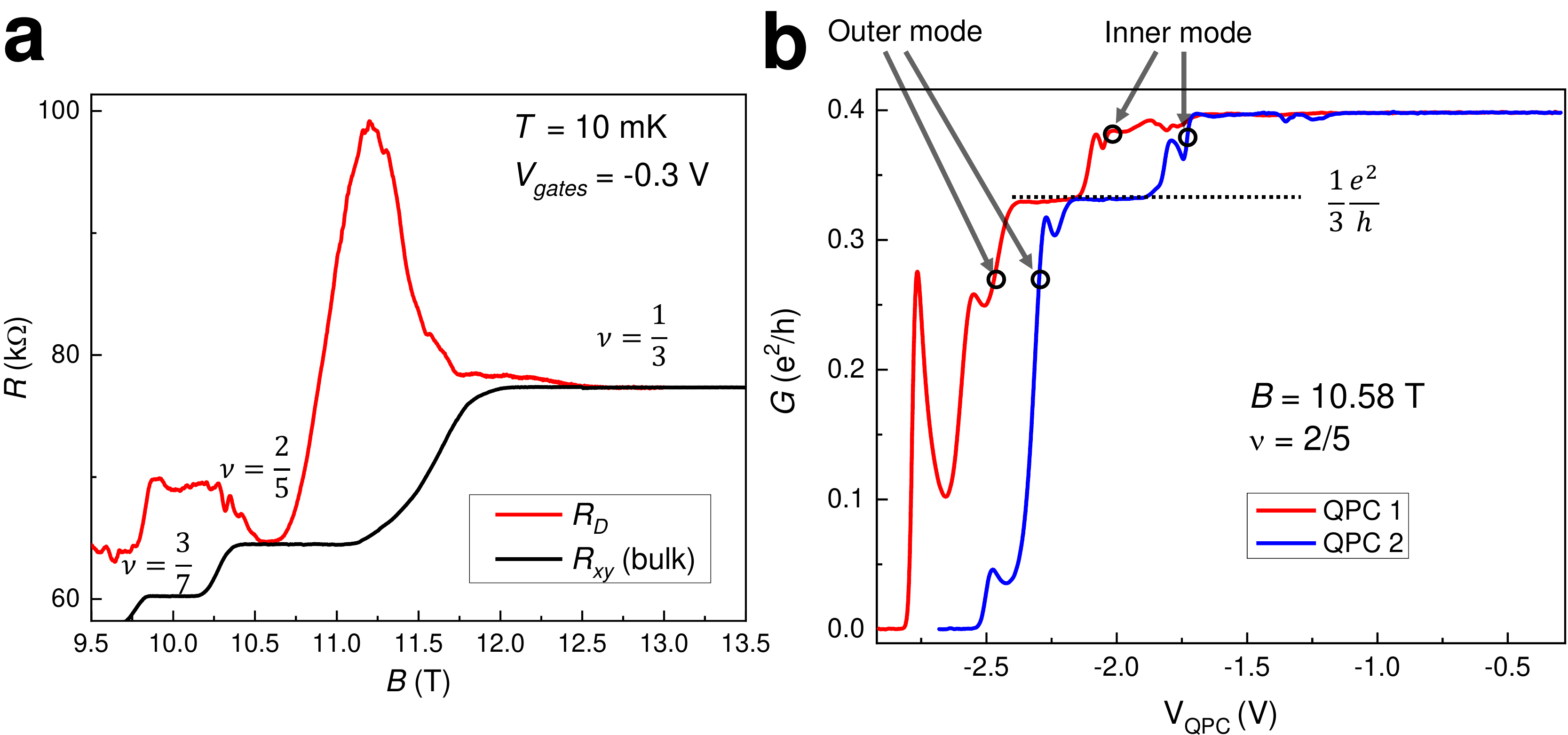}
\caption{\label{RD}a) Diagonal resistance $R_D$ measured across the interferometer with the 2DEG under the gates just depleted so that current flows through the device (red), compared to bulk transport $R_{xy}$ (black). While $R_{D}$ is generally larger the $R_{xy}$ due to scattering induced within the device, at $\nu = 1/3$ and $\nu = 2/5$ there are regions where $R_D$ reaches the bulk value of $R_{xy}$, indicating full transmission of the edge states with minimal scattering through the middle of the interferometer. b) Conductance measured across the interferometer as a function of the QPC voltage for each individual QPC at $\nu = \frac{2}{5}$, $B = 10.58$ T. The conductance starts at the full value of $G = \frac{2}{5} \frac{e^2}{h}$, and decreases as more negative voltage is applied to bring the edges together and induce backscattering. A clear intermediate plateau at $G = \frac{1}{3}\frac{e^2}{h}$ is visible, indicating full reflection of the inner edge while the outer edge state is fully transmitted. Approximate operating points for interference of the inner and outer modes are indicated with circles and arrows. }
\end{figure}

In Fig. \ref{RD}a we show measurements of the bulk Hall resistance $R_{xy}$ (black) at high magnetic field in the fractional quantum Hall regime, measured in a region of 2DEG away from the interferometer. Prominent resistance plateaus occur at $\nu = 1/3$ and $\nu = 2/5$. Measurement of the diagonal resistance across the device $R_{D}$ is displayed in red, with a small negative voltage of -0.3V on the QPCs and side gates in order to deplete the electrons under the gates, but not induce significant backscattering in the QPCs. Given any backscattering in the device, $R_D$ will generally be higher than $R_{xy}$, but note there is a wide range of magnetic field at $\nu=1/3$ where $R_{D} \approx 3 \frac{h}{e^2}$ while the $R_{xy}$ remains quantized, indicating nearly full transmission of the edge state through the device. At $\nu = 2/5$, there is a range of magnetic field approximately 200mT wide where $R_{D} \approx 2.5 \frac{h}{e^2}$, indicating that both edge modes are nearly fully transmitted through the device with minimal conduction through the bulk. 

In Fig. \ref{RD}b we show conductance versus QPC gate voltage for the two QPCs individually at $\nu = 2/5$. There is a wide primary plateau with $G \approx 0.4 e^2/h$ where both edge modes are nearly fully transmitted through the device. As $V_{QPC}$ is made more negative, conductance decreases and then reaches a second, intermediate plateau at $G\approx \frac{1}{3} \frac{e^2}{h}$, indicating that the inner edge state is fully reflected while the outer one (which carries conductance $\frac{1}{3} \frac{e^2}{h}$) is fully transmitted. Beyond this second plateau, the outer edge state starts to be reflected, and conductance drops until reaching zero. Tuning the QPCs to values between the first and second plateaus corresponds to partial reflection of the inner mode, while tuning conductance between the second plateau and zero conductance corresponds to partially reflecting the outer edge mode while the inner one is fully reflected; thus, it is possible to set $V_{QPC}$ on both QPCs to select which edge mode is interfered. This {\it in-situ} tuning of individual edge mode transmission is critical to our experiment. Approximate gate operating points for interference of the inner mode and outer mode are indicated with arrows in the figure (these points are not exact since there is a small amount of cross coupling between the gates). 

\section{Interference of Outer Mode}

\begin{figure}[h]
\def\ffile{Outer_Mode_V2.pdf}
\centering
\includegraphics[width=1.0\linewidth]{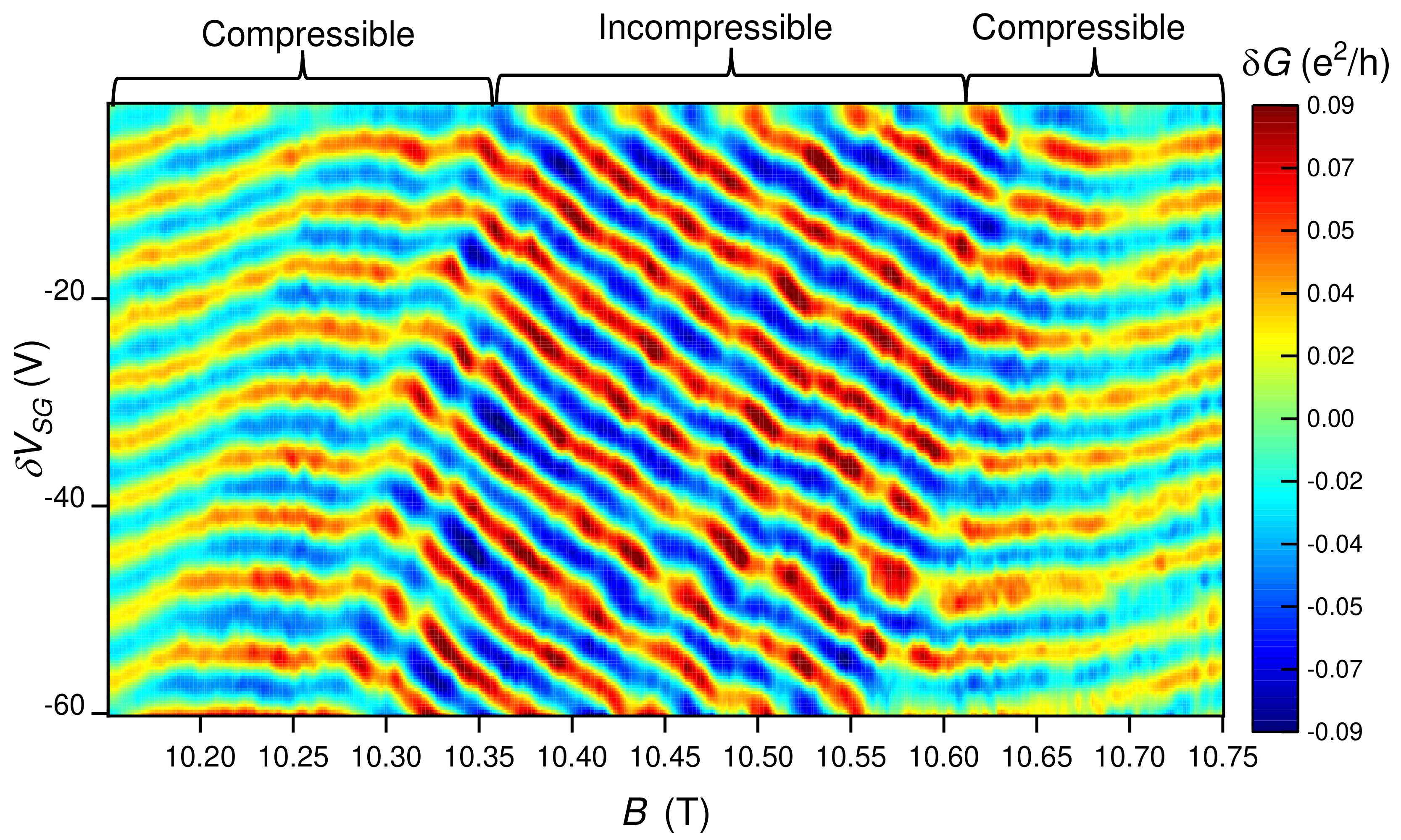}
\caption{\label{Outer} Conductance oscillations versus $B$ and $\delta V_{SG}$ for the outer mode at $\nu = \frac{2}{5}$. The oscillations resemble those observed at $\nu = 1/3$, with a central incompressible region with negatively sloped lines of constant phase, and the lines of constant phase becoming nearly flat at higher and lower fields when the density of states in the center of the device is high. Discrete jumps in phase due to changing quasiparticle number are less clearly identifiable than at $\nu = 1/3$.}
\end{figure}

We begin our analysis with examination of interference at $\nu=2/5$ by considering the outer edge mode. Fig. \ref{Outer} shows conductance oscillations with the QPCs tuned to weakly backscatter the outer edge mode (with $\approx 85 \%$ QPC transmission; approximate operating points indicated in Fig. \ref{RD}b). There is a central region where lines of constant phase follow a negative slope, indicative of Aharonov-Bohm interference, with lines of constant phase becoming nearly flat at higher and lower field. This behavior is very similar to interference observed at $\nu = 1/3$, where the transitions from AB interference to nearly flat lines of constant phase is caused by transitions from an incompressible to a compressible bulk; in the compressible regions quasiparticles/quasiholes are removed/created with $\Phi_0$ period. The fact that the outer edge mode interference at $\nu=2/5$ exhibits a similar phenomenon indicates that this edge mode has the same properties as the single edge state at $\nu = 1/3$, in agreement with theoretical expectations \cite{Stern2008, Halperin2011} and recent experiments probing the outer edge at $\nu = 2/5$ \cite{Glidic2023, Ruelle2023, Lee2022, Kundu2023}. In the composite fermion model \cite{Jain1989, Jain1990, JainBook}, this similarity can be understood from the fact that both the $\nu = 1/3$ edge state and the outer edge state at $\nu = 2/5$ are generated from the lowest Lambda level, and therefore are expected to have the same properties. While the localized quasiparticles of the bulk 2/5 state have charge $e/5$, as discussed in Ref. \cite{Halperin2011}, when the inner mode is fully reflected (resulting in a filling factor $\nu \leq 1/3$ in the QPCs), it is expected that the overall charge inside the device should be quantized in units of $e/3$ \cite{Stern2008, Halperin2011}, and thus the the relevant $\theta_a$ in Eqn. \ref{theta_AB} is $\frac{2\pi}{3}$ since the localized inner puddle with $\Delta \nu = \frac{1}{15}$ can be considered to be composed of charge $e/3$ anyon quasiparticles.

While the outer edge state at $\nu=2/5$ shares properties with the $\nu = 1/3$ edge state, the properties of the bulk are different, which leads to some differences in the interference behavior. The energy gap of the $\nu=2/5$ state is significantly smaller than $\nu = 1/3$, leading to a narrower region of magnetic field where the device exhibits negatively-sloped incompressible regime oscillations \cite{Rosenow2020}. We have measured an activated transport energy gap $\Delta^{2/5}$=3.6 K at $\nu = 2/5$; based on the model of Rosenow \& Stern \cite{Rosenow2020}, the range of field where the bulk is incompressible should be given by $\Delta B_{inc} = \frac{\Delta \Phi_0 C}{\nu e^* e^2}$, with $C \approx 4.4 \times 10^{-3}$ $\frac{\mathrm{F}}{\mathrm{m}^2}$ the combined capacitance per unit area of the screening layers and $\Delta =3.6$ K the energy gap. Using measured parameters for our device, this yields an expected incompressible region of $\approx 430$ mT, in reasonable agreement with our observed region of $\approx 300$ mT where the outer mode exhibits negatively slope oscillations.

There are additional important differences in behavior between the single edge mode at $\nu = 1/3$ and the outer mode at $\nu = 2/5$. At $\nu = 2/5$ even when the inner puddle is gapped and incompressible, the edge of the inner puddle remains gapless. Thus edge quasiparticles can be added to the inner edge mode when the magnetic field and gate voltage are varied. Due to disorder, even in the incompressible inner puddle there will be a finite density of localized $e/5$ quasiparticle states such that when $B$ and $\delta V_{SG}$ are varied, these localized states are occupied and the overall number of $e/3$ quasiparticles in the inner puddle may change. While at $\nu = 1/3$ transitions in the number of localized quasiparticles were seen to occur along nearly straight positively-sloped lines in the $B-\delta V_{SG}$ plane \cite{Nakamura2020, Nakamura2022}, at $\nu=2/5$ since quasiparticles can be added both to the interior and to the edge of the inner puddle, the dependence of $N_{qp}$ on $B$ and $\delta V_{SG}$ will be more complicated. Fig. \ref{Outer} does show modulations in the incompressible region which likely correspond to creation of localized quasiparticles and modulations of $\theta$ via $\theta_a$, although it might be considered that the $e/3$ quasiparticles have ``lost their identity'' \cite{Halperin2011} in forming the inner condensate.

\begin{figure}[h]
\def\ffile{Up_to_1_3_v2.pdf}
\centering
\includegraphics[width=1.0\linewidth]{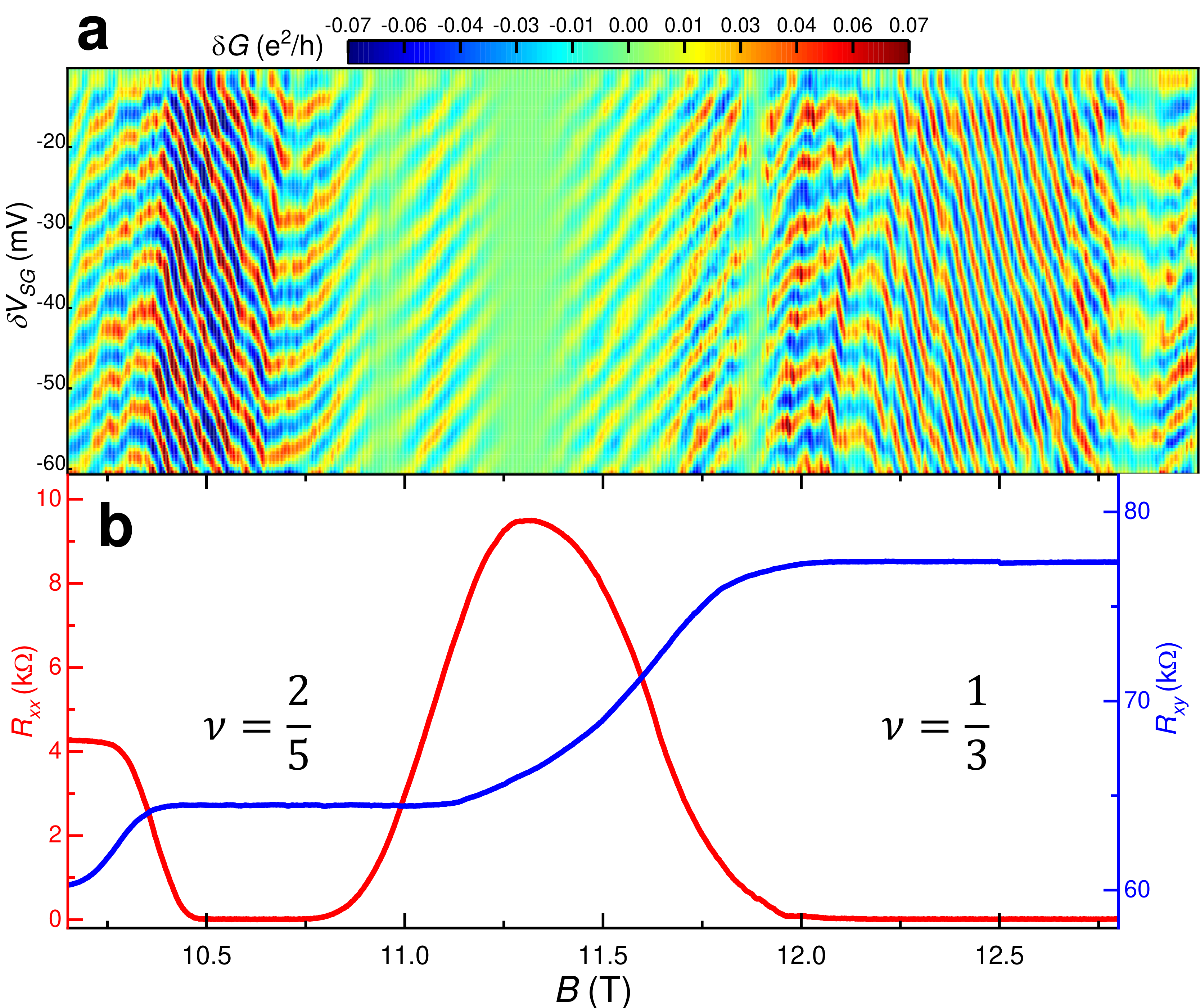}
\caption{\label{Outer_1_3} Interference of the outer edge mode spanning filling factor $\nu=2/5$ to $\nu=1/3$ (upper panel a) compared to bulk transport $R_{xx}$ and $R_{xy}$ (lower panel b). The conductance oscillations are continuous from $\nu = 2/5$ to $\nu = 1/3$. The QPCs are continuously adjusted as magnetic field is varied in order to maintain weak backscattering. There is is a clear suppression of interference amplitude in the region where $R_{xx}$ has a peak between $\nu = 2/5$ and $\nu = 1/3$, suggesting that high conductivity in the bulk suppresses interference. Interference at $\nu = 1/3$ is consistent with previous works \cite{Nakamura2019, Nakamura2020, Nakamura2022}, with a wide incompressible region with a small number of discrete phase jumps due to anyonic statistics, with the lines of constant phase flatting out at high and low field when the bulk becomes compressible.}
\end{figure}

To further investigate the correspondence between the outer mode at $\nu = 2/5$ and the edge mode at $\nu = 1/3$, we monitor interference over the entire range of magnetic field from $\nu=2/5$ to $\nu=1/3$ as shown in Fig. \ref{Outer_1_3}a. Because the QPC voltage required to weakly backscatter the outer mode at 2/5 is slightly more negative than the voltage required to backscatter the single edge state at $\nu = 1/3$ (likely due to the outer edge state moving inward as field is increased and the inner edge state vanishes), for this measurement we continuously adjust the QPC voltages as $B$ is varied in order to maintain interference. Bulk magnetotransport is shown in Fig. \ref{Outer_1_3}b. The oscillations are continuous from $\nu = 2/5$ to $\nu = 1/3$, verifying that the edge state persists in both quantum Hall states. Interestingly, while at both $\nu=1/3$ and $\nu=2/5$ the lines of constant phase become nearly flat above and below the incompressible region, in the transition region between $\nu = 2/5$ and $\nu = 1/3$ the lines adopt a shallow positive slope and interference is greatly suppressed. A possible explanation for this is that in this region the area enclosed by the edge state shrinks, since it transitions from being an outer edge state to being the only edge state in the system, which might cause it to move slightly inward. In this transition region the interference amplitude is dramatically suppressed. This may occur because in this region, the bulk of the 2DEG is highly conductive (as can be seen from the large peak in $R_{xx}$), leading significant smearing of the localized quasiparticle number as the transitions are broadened \cite{Feldman2022}, as well as a less well defined interference path. 

At $\nu = 1/3$, we observe behavior consistent with our previous experiments: there is a wide incompressible region where the device exhibits negatively-sloped AB interference with a few discrete jumps in phase, with lines of constant phase flattening out at lower and higher magnetic fields when the bulk becomes compressible. More data at $\nu=1/3$ is shown in Supp. Fig. 8 and discussed in Supp. Section 8. The width of the incompressible region is $\approx 600$ mT, approximately a factor of two larger than at $\nu = 2/5$, reflecting the larger energy gap of the state. 

\section{Interference of Inner Mode}

\begin{figure}[h]
\def\ffile{Inner_Wide_v4.pdf}
\centering
\includegraphics[width=1.0\linewidth]{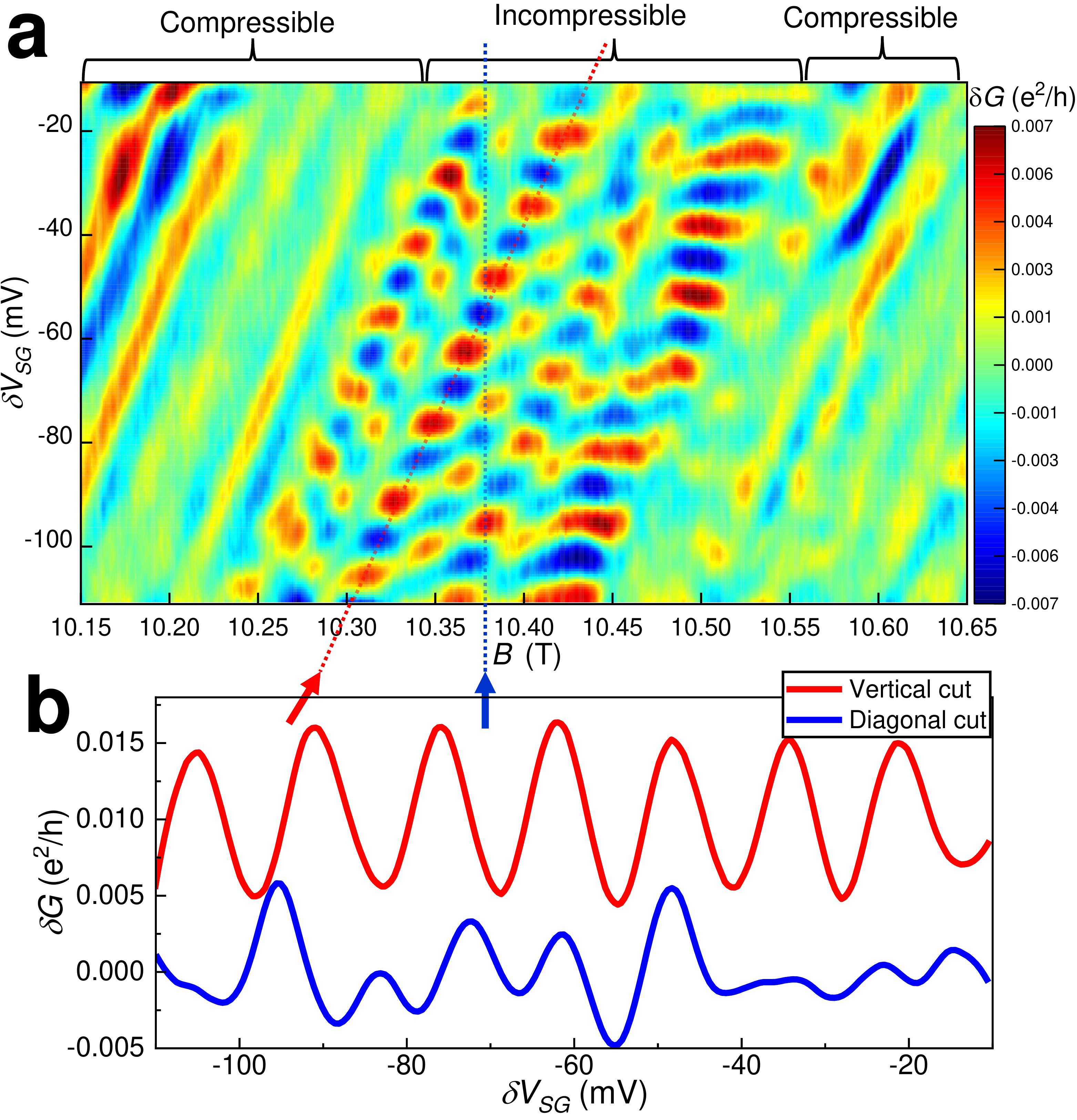}
\caption{\label{Inner_Wide}a) Conductance verseus $B$ and $\delta V_{SG}$ for the inner mode at $\nu = \frac{2}{5}$. There is a central region where a checkerboard-pattern forms with discrete jumps in the oscillation pattern indicative of anyonic statistics. At higher and lower magnetic fields where the bulk becomes compressible, the interference oscillations disappear, and weak positively sloped oscillations occur, which are most likely explained by conduction through the bulk rather than an interference process. b) Vertical cut of conductance versus $\delta V_{sg}$ (blue) in the incompressible regime. Since this cut intersects several of the discrete jumps in phase, the behavior is non-sinusoidal. On diagonal cuts parallel to (but in between) the discrete jumps in phase (red), the quasiparticle number is fixed, so the conductance oscillates sinusoidally due to the continuously varying Aharonov-Bohm phase. Red and blue dashed lines in a) in a indicate where each cut is taken. }
\end{figure}

\begin{figure}[t]
\def\ffile{Inner_Mode_jumps.pdf}
\centering
\includegraphics[width=1.0\linewidth]{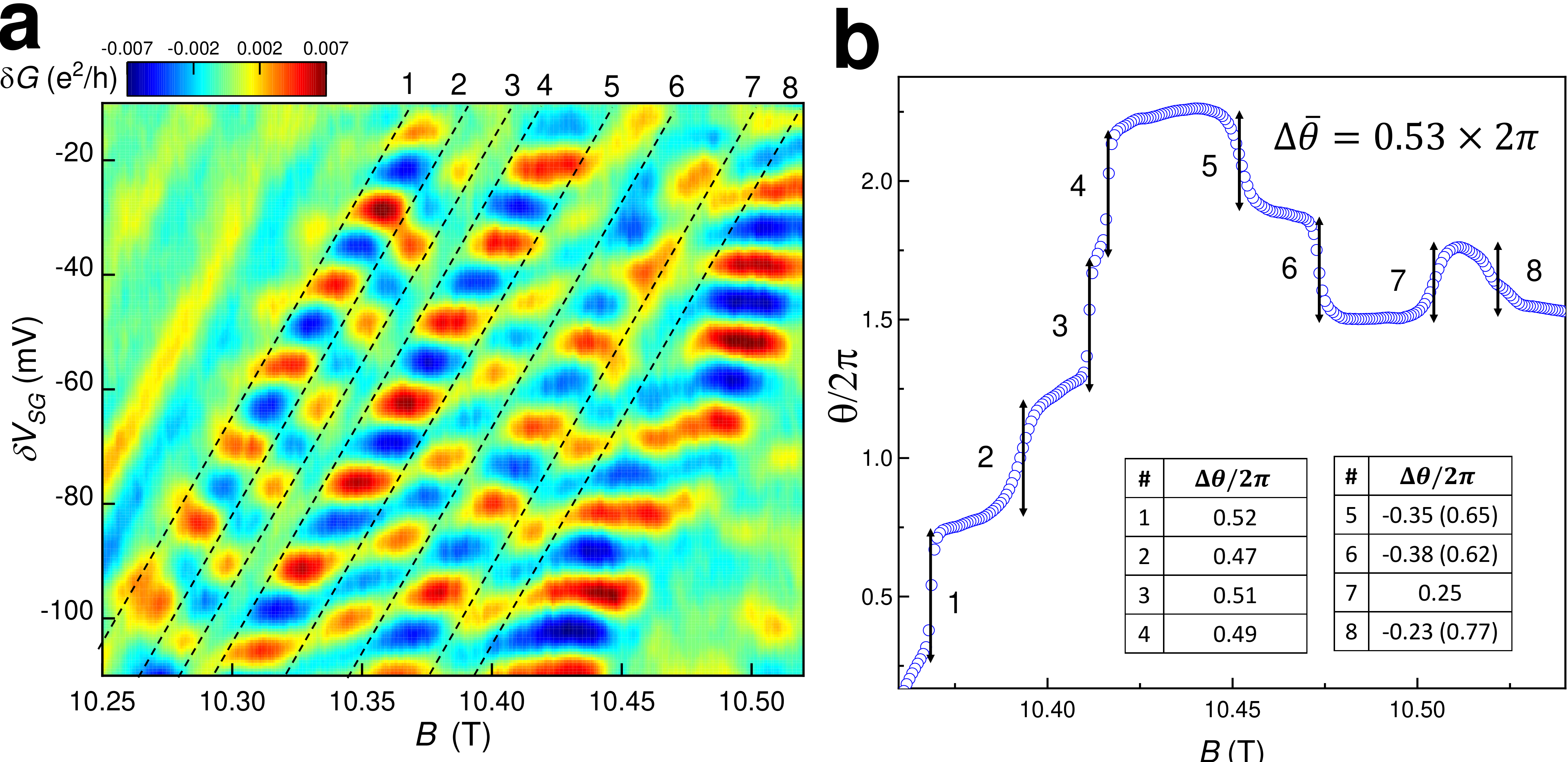}
\caption{\label{Inner}a) Conductance versus side gate voltage and magnetic field for the inner mode at $\nu= \frac{2}{5}$ (zoomed in view on the incompressible region from Fig. \ref{Inner_Wide}a) with dashed lines indicating the positions of discrete jumps in phase. b) Phase extracted via Fourier transform versus magnetic field. Discrete steps in the phase correspond to the discrete jumps visible in a. The values of each phase jump (calculated from difference of average $\theta$ on each plateau between the jumps) is indicated in the table above the plot. }
\end{figure}

In Fig. \ref{Inner_Wide}a we show conductance as a function of side gate voltage variation $\delta V_{SG}$ (this side gate voltage variation is relative to -1.8 V and applied symmetrically to both side gates) and magnetic field $B$ with the QPCs tuned to weak backscattering of the inner mode (both QPCs initially tuned to approximately 75\% transmission of the inner mode; approximate operation points shown in Fig. \ref{RD}b). In contrast to the relatively simple behavior of $\nu=1$ where clear lines of constant phase follow a negative slope, for $\nu = 2/5$ the behavior is not a simple sinusoidal function of $B$ and $\delta V_{SG}$. Instead, there are many discrete jumps in phase (indicated with dashed lines in Fig. \ref{Inner}a) which follow a positive slope in the $B$ and $\delta V_{SG}$ plane. These discrete jumps in phase create a checkerboard-like pattern in the conductance. It is also noteworthy that the amplitude is small, approximately an order of magnitude smaller than interference of the outer mode at $\nu=2/5$ or at $\nu = 1/3$; this hints that interference of the edge mode is significantly more prone to decoherence \cite{Feldman2022}. Nevertheless, the interference pattern is temporally stable and repeatable in subsequent scans (see Supplemental Section 3 and Supp. Fig 3). Vertical cuts of conductance versus $\delta V_{SG}$ at fixed $B$ (indicated with the blue arrow in Fig. \ref{Inner_Wide}a, plotted in blue in Fig. \ref{Inner_Wide}b) show non-sinusoidal behavior due to the discrete jumps in phase. In stark contrast, line cuts parallel to (but in between) the discrete jumps in phase (indicated by the red arrow in \ref{Inner_Wide}a, plotted in red in Fig. \ref{Inner}b) do show clear sinusoidal oscillations, indicating that along these contours the number of localized quasiparticles is fixed and the phase variation is only due to the continously-varying Aharonov-Bohm phase in Eqn. \ref{theta_AB}. 

\section{FRACTIONAL CHARGE}
In regions between the discrete jumps, where the phase varies only due to the Aharonov-Bohm contribution, the gate voltage oscillation period is $\approx 14$ mV. With the gate lever arm $\frac{\partial{\Bar{A}}}{\partial V_{SG}} = 0.11$ $ \mu$m$^2$V$^{-1}$, this implies a flux period of $\approx 4 \Phi_0$ if bulk-edge coupling is neglected. To quantitatively extract the effective quasiparticle charge, we must account for both the change in area due to direct coupling of the gate to the edge and the effect of bulk-edge coupling from Eqn. \ref{theta_BE}. This analysis yields the following equation for the gate voltage oscillation period $\Delta V_{SG}$ when the bulk is incompressible:

\begin{equation}
    \label{delta_Vg}
    \Delta V_{SG} = \frac{\Phi_0}{B e^*}(\frac{\partial \Bar{A}}{\partial V_{SG}} +\frac{\kappa \alpha _{bulk} \Phi_0}{\Delta \nu B})^{-1}
\end{equation}
where $\alpha_{bulk} = \frac{\partial \Bar{q_b}}{\partial V_{SG}}$ and $\kappa = -\frac{\delta {q_i}}{\delta {q_b}}$.
$\kappa = 0.17$ is extracted for our device from finite bias conductance measurements (see Supp. Section 6, Supp. Section 7, and Supp. Fig. 6, and Refs. \cite{Roosli2020b, Sivan2016a, Frigeri2019, Beenakker1991, Kouwenhoven1999, Wen1991a, Smits2013, Fujisawa2022, Chklovskii1993, Sahasrabudhe2018}) and $\alpha _{bulk}=0.06$ mV$^{-1}$ is determined with the aid of zero magnetic field Coulomb blockade spectroscopy (see Supp. Section 2).
Applied to our device with $\Delta V_{SG} = 14$ mV, Eqn. 3 yields $e^* = 0.17 \pm0.02$ for the inner mode (here uncertainty is estimated from the FWHM of the Fourier transform peak), close to the theoretical value of $e/5$. This confirms the interference of fractionally charged quasiparticles at $\nu = 2/5$, and supports previous measurements of fractional charge at $\nu = 2/5$ \cite{Reznikov1999}.

It is noteworthy that in the regions between the discrete jumps where the quasiparticle number is constant, the phase evolves with gate voltage (as shown in Fig. \ref{Inner_Wide}b), but appears to be nearly independent of magnetic field. This contrasts with the naive expectations of Eqn. \ref{theta_AB}, which would imply a phase evolution with $B$ and $\frac{\Phi_0}{e^*}$ oscillation period. On the other hand, Eqn. \ref{theta_BE} implies that in the incompressible region where $N_{qp}$ is fixed there will be a crossover from negatively-sloped AB oscillations to positively sloped oscillations at a critical value of $\kappa = \frac{\Delta \nu}{\nu_{in}} = \frac{(1/15)}{(2/5)} = \frac{1}{6}$; this $\kappa$ corresponds to the transition from the Aharonov-Bohm to the Coulomb-dominated regime for $\nu=2/5$, and also applies to the regime of a compressible bulk. Note that a significantly larger bulk-edge coupling parameter is required for integer quantum Hall states states to transition from Aharonov-Bohm to Coulomb dominated interference, with the critical value being $\kappa = 0.5$. 

This critical value of $\frac{1}{6}$ for $\nu = 2/5$ is very close to the value of $\kappa = 0.17$ from finite-bias measurements; thus, bulk-edge coupling does account for the weak $B$ dependence in the incompressible regions between the discrete jumps. It is interesting to note that while the intrinsic bulk-edge coupling, quantified by $\kappa$, is not particularly strong, its impact at $\nu=2/5$ is significant.  Simulations of interference for the inner mode illustrating this effect for different values of $\kappa$ are shown in Supp. Fig. 3 and discussed in Supp. Section 3.

Next we discuss in detail the discrete jumps in phase, which are indicated with dashed lines in Fig. \ref{Inner}a. Qualitatively, these discrete jumps in phase are more dramatic features than the anyonic phase jumps which have been previously observed at $\nu = 1/3$ \cite{Nakamura2020, Nakamura2022}, consistent with the larger magnitude the $\theta _a$ at $\nu=2/5$ compared to $\nu=1/3$. The positive slope of the transition lines (dashed line in Fig. \ref{Inner}a) in the $B-V_{SG}$ plane is consistent with previous observations of anyonic statistics for the $\nu = 1/3$ state. If the side gates coupled only to the edge states, then the gates would not affect the localized quasiparticle number, and the discrete jumps would be vertical (only occurring as a function of $B$). However, real devices have some coupling of the side gate to the bulk, leading to an excess bulk charge $\delta q_b = \frac{\nu \Bar{A} \delta B}{\Phi_0}-\alpha _{bulk}\delta V_{SG}+e^* N_{qp}$, causing the contours of fixed bulk charge (across which quasiparticle transitions happen) to have slope $\frac{d V_{SG}}{dB} = \frac{\nu \Bar{A}}{\Phi_0 \alpha_{bulk}} \approx 0.64$ mV/mT. The observed slope is $\approx 0.8$ mV/mT, close to the value calculated from these simple charge balance considerations.   

In Fig. \ref{Inner}c we plot the interference phase versus magnetic field (the phase is extracted via Fourier transform diagonal cuts of conductance (see Supp. Section 4 and Supp. Fig. 4). Steps in the phase occur corresponding to the discrete jumps indicated in Fig. \ref{Inner}a. Since there is weak magnetic field dependence in the regions between the jumps, the Aharonov-Bohm contribution to the phase is small, so the steps in phase correspond primarily to the transitions in quasiparticle number (this is supported by the fact that the phase is nearly flat in the regions between the discrete jumps, even without subtracting an AB component). The values of the phase jumps (calculated as the difference in average phase on the plateaus on either side of the jump) are listed in the table inset in Fig. \ref{Inner}b. Interestingly, there are both positive and negative jumps in phase, in contrast to the naive expectation of $\Delta \theta = -\theta_a = \frac{4\pi}{5}$ in the absence of bulk-edge coupling (the negative sign comes from that the fact that increasing $B$ is expected to remove quasiparticles or equivalently add quasiholes). This indicates that bulk-edge coupling, even at $\kappa \simeq 0.17$, is indeed an important factor.

\section{Quasiparticle statistics}

With bulk-edge coupling, the change in interference phase when a quasiparticle is removed based on Eqn. \ref{theta_BE} is $\Delta \theta =-\theta _a+\kappa \frac{(e^*) ^2}{\Delta \nu}$; for the inner mode of $\nu = 2/5$ with $e^* = 1/5$, $\Delta \nu = 2/5-1/3 = 1/15$, and $\theta _a=\frac{-4\pi}{5}$, this yields 

\begin{equation}
\label{delta_theta}
\frac{\Delta \theta}{2\pi}=-\frac{\theta_a}{2\pi}+\kappa \frac{(e^*)^2}{\Delta \nu}=(\frac{2}{5}+\kappa \frac{3}{5})
\end{equation}

The factor of $\frac{1}{\Delta\nu}$ makes interference at $\nu=2/5$ significantly more sensitive to bulk-edge coupling than at $\nu = 1/3$. This implies that a moderate bulk-edge coupling parameter of $\kappa = \frac{1}{6}$ (the same critical value as for the Aharonov-Bohm slope) can push $\Delta \theta$ to the transition point of $0.5\times 2\pi$, where the phase jumps cross over from positive to negative. Thus, our observation observation of a mix of positive and negative phase jumps may be explained by an average bulk-edge coupling parameter of $\kappa \approx \frac{1}{6}\approx 0.167$, with small variations in the exact coupling of individual localized quasiparticle states to the edge leading to some quasiparticle transitions giving a positive change in phase and some negative (with the phase defined from $-\pi$ to $+\pi$). On the other hand, based on Eqn. \ref{delta_theta}, it is more convenient to define jumps in phase from 0 to $2\pi$ in order to quantitatively account for the effect of bulk-edge coupling. The values of the discrete jumps are listed in the inset to Fig. \ref{Inner}b; for the negative jumps, the value shifted into the range 0 to $2\pi$ is shown in parenthesis. We calculate the average $\Delta \bar{\theta}$ using the values from 0 to $2\pi$, which yields $\Delta \bar{\theta}=(0.53\pm0.05) \times 2\pi$ (note that the standard deviation is $0.15 \times 2\pi$, and the uncertainty is estimated by the standard error). This average value is somewhat higher than the ideal value of $\Delta \theta = -\theta_a = \frac{4\pi}{5}$, but as indicated by Eqn. \ref{delta_theta}, finite bulk-edge coupling tends to increase $\Delta \theta$, so an experimental value somewhat higher than the ideal value is expected.


Based on Eqn. \ref{delta_theta}, the anyonic phase can be extracted from the phase jumps as $\theta _a = -\bar{\Delta \theta}+2\pi \kappa \frac{(e^*)^2}{\Delta \nu}$. Using the value of $\kappa = 0.17$ extracted from finite bias measurements yields $\theta _a = -(0.43 \pm 0.05) \times 2\pi$, in good agreement with the value of $\theta_a = -\frac{4\pi}{5}$ predicted from theory \cite{Halperin1984, Su1986, Stern2008, Halperin2011} and numerical work \cite{Jeon2003a, Jeon2004}. Thus, our experiment confirms the theoretical prediction of anyonic braiding statistics at $\nu = 2/5$ with statistical angle $\theta_a = -\frac{4\pi}{5}$. This is the first quantitative experimental determination of the anyonic braiding phase at the hierarchy $\nu=2/5$ state.

\section{Behavior in the compressible regime}

As can be seen in Fig. \ref{Inner_Wide}a, there is a limited range of field where the checkerboard pattern of interference created by discrete jumps in phase is visible; above and below this region there are weak oscillations with a large magnetic field period (corresponding to $\approx 4 \Phi_0$) and a positive slope in the $B-\Delta V_{SG}$ plane. The values of magnetic field where these transitions occur are similar to those where the bulk transitions from incompressible to compressible when interfering the outer mode, as seen in Fig. \ref{Outer}, suggesting that the transition in behavior also occurs due to the bulk becoming compressible. However, the positive-slope oscillations in the compressible regions are not consistent with interference in the fully compressible regime. With a fully compressible bulk, an $e/5$ quasiparticle/quasihole would be added/removed with each $\frac{\Phi_0}{2}$ increase in flux in order to keep the density of the 2DEG constant. This would result in oscillations with $\Phi_0$ magnetic field period based on Eqn. \ref{theta_AB} (the lines of constant phase could be positive or negative depending on the degree of bulk-edge coupling, but this $\Phi_0$ period would remain the same). Therefore, these positively slope lines are unlikely to be from an interference process, and are more likely to be associated with conduction through the bulk of the interferometer. Additionally, the slope of these oscillations is $\approx 0.8$ mV/mT, close to the value of the slope of the quasiparticle transitions seen in Fig. \ref{Inner}a. This supports the possibility that these oscillations actually occur due to a resonant tunneling process through the middle of the interferometer rather than true interference process where tunneling only occurs at the QPCs. This bulk contribution to conductance is consistent with the observation from Fig. \ref{RD}a that conductance through the device is only quantized in a very narrow window of magnetic field even when minimal backscattering is induced in the QPCs.

The absence of compressible-regime interference for the inner mode is interesting, given that the outer mode does show interference in this regime. A possible explanation for this is thermal smearing of the quasiparticle number, leading to topological dephasing \cite{Park2015}. A rapid suppression of amplitude with temperature was previously observed at $\nu = 1/3$ in the compressible regime due to this mechanism \cite{Nakamura2020} (note also that for the outer mode at $\nu=2/5$, the amplitude is significantly suppressed in the compressible regimes as can be seen in Fig. \ref{Outer}). For the inner mode at $\nu=2/5$, the relevant energy scale for confining quasiparticles should be much smaller due to the small quasiparticle charge $e/5$, making it plausible that the oscillations are so strongly thermally smeared that they are not measurable. Even in the {\it incompressible} regime, we find a small temperature decay scale $T_0 = 26$ mK (see Supp. Section 8 and Supp. Fig. 7), so it appears reasonable that interference is unmeasurable in the compressible regime where the temperature decay scale will be even smaller. This is supported by simulations (Supp. Fig. 5).

\section{Conclusions}
We have measured conductance oscillations due to interference of the inner and outer edge modes at $\nu = 2/5$. The outer mode exhibits behavior similar to $\nu = 1/3$, supporting the expectation that the outer edge state at $\nu=2/5$ has the same properties as the single edge state at $\nu = 1/3$, although the smaller energy gap results in a narrower range of magnetic field over which the bulk is incompressible. The inner edge exhibits oscillations with a gate voltage period consistent with an interfering charge $e^* = 0.17\pm 0.02$, close to the theoretically predicted value of $\frac{1}{5}$. Discrete jumps in phase with average value $\Bar{\Delta \theta} = 0.53$ are observed; after taking into account the impact of bulk-edge coupling, we extract an anyonic braiding phase $\theta _a = (-0.43\pm 0.05) \times 2\pi$, close to the theoretically anticipated value $\theta_{a}=-\frac{4\pi}{5}$. These measurements give experimental support to the theoretical prediction of anyonic quasiparticles at the $\nu = 2/5$ state, and demonstrate that Fabry-P{\'e}rot interferometry can be extended to fractional quantum Hall states with multiple edge modes to make quantitative measurements.

\section{Acknowledgements}
This work was supported by the U.S. Department of Energy, Office of Science, Office of Basic Energy Sciences, under Award No. DE-SC0020138. We wish to thank Bert Halperin and Dima Feldman for several insightful discussions and suggestions that improved our manuscript.

\bibliography{Interferometer_Bib}

\end{document}


\renewcommand{\figurename}{SUPP. FIG.}

\title{Supplementary Material for ``Fabry-P{\'e}rot interferometry at the $\nu$ = 2/5 fractional quantum Hall state''}

\maketitle

\section{Supplementary Section 1: Heterostructure design}
\begin{figure}[h]
\def\ffile{Structure.pdf}
\centering
\includegraphics[width=1.0\linewidth]{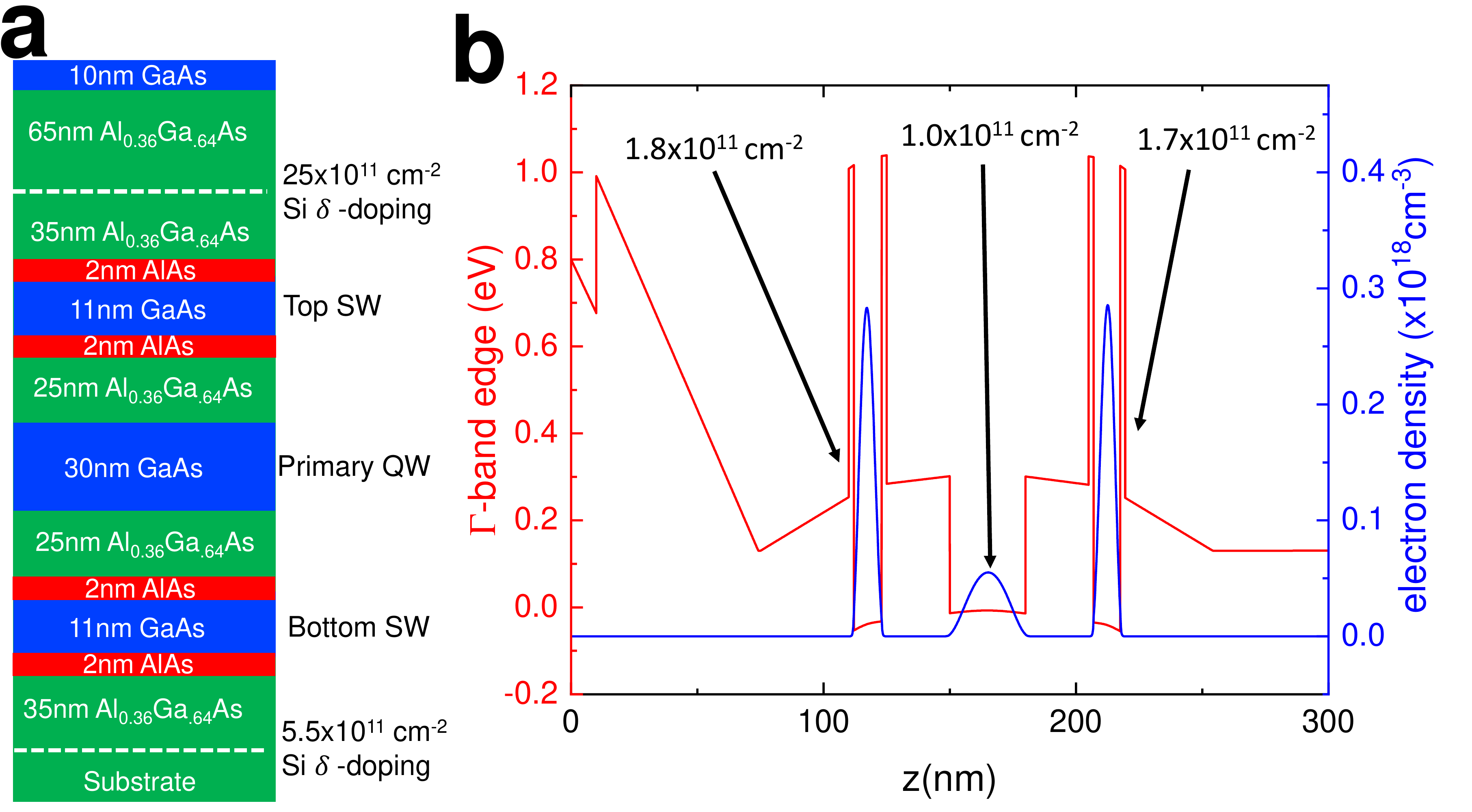}
\caption{\label{Structure}a) Layer stack for the GaAs/AlGaAs heterostructure used for these experiments. The structure includes a primary GaAs quantum well, and additional screening wells above and below the main well. b) Simulation of $\Gamma$-band edge and electron density in the heterostructure using the Nextnano simulation package \cite{Birner:2007}. }
\end{figure}

The heterostructure used in this experiment uses the screening well design \cite{Nakamura2019}, similar to that used in our previous interferometer experiments probing the $\nu = 1/3$ state \cite{Nakamura2019, Nakamura2020, Nakamura2022}. The layer stack of the structure is shown in Supp. Fig. \ref{Structure}a. The screening wells screen the long-range Coulomb interaction in the device, effectively lowering the charging energy and reducing the bulk-edge interaction; this is necessary in order for anyonic statistics to be observable in mesoscopic devices \cite{Rosenow2007, Halperin2011}. Compared to previous works which detected evidence for anyonic statistics at $\nu = 1/3$ \cite{Nakamura2020, Nakamura2022}, the current heterostructure has a narrower screening wells. This results in more charge transfer to the main quantum well, which increases the electron density and thus increases the energy gap and robustness of the the $\nu = 2/5$ fractional quantum Hall state. Supp. Fig. \ref{Structure}b shows a simulation of the heterostructure using the Nextnano software package \cite{Birner:2007}. Additionally, the setback of the donors from the screening wells is reduced so that the electron density in the screening wells is increased compared to Refs. \cite{Nakamura2020, Nakamura2022}, ensuring that the screening well density is significantly higher than the main quantum well density, in order to maintain strong screening of the bulk-edge interaction.

\section{Supplementary Section 2: Extraction of device parameters}

\begin{figure}[h]
\def\ffile{Integer_Periods.pdf}
\centering
\includegraphics[width=1.0\linewidth]{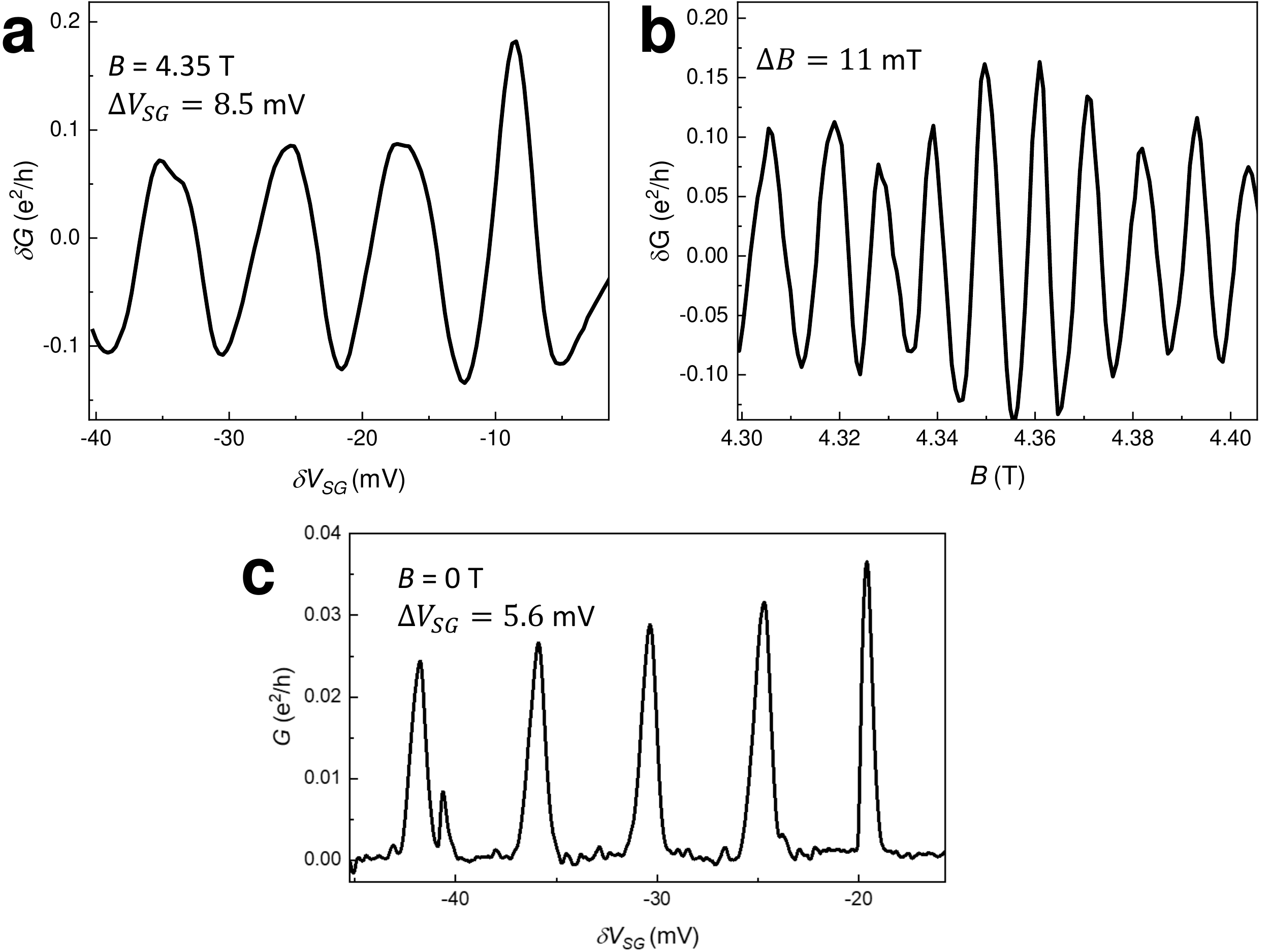}
\caption{\label{Periods} Conductance oscillations at $\nu=1$ in Device A. a) Line cut of conductance versus gate voltage at $\nu = 1$, $B = 4.35$ T. b) Line cut of conductance versus magnetic field at $\nu = 1$. c) Conductance versus gate voltage in the Coulomb blockade regime at $B = 0$.}
\end{figure}

In order to interpret interference as a function of gate voltage and magnetic field, it is necessary to understand the coupling constants and effective area of the interferometer. We extract the lever arm relating change in area to change in gate voltage from the gate voltage oscillation period at $\nu = 1$, shown in Fig. \ref{Periods}a. The oscillation period is $\Delta V_{SG} = 8.5$ mV, yielding $\frac{\partial \Bar{A}}{\partial V_{SG}} = \frac{\Phi_0}{B \Delta V_{SG}} \approx 0.11$ $\mu$m$^2$V$^{-1}$.

The effective interferometer area $\Bar{A}$ (not the small variations due to bulk-edge coupling) can be extracted from the magnetic field oscillation period. This yields $\Bar{A} = \frac{\Phi_0}{\Delta B} = 0.38$ $\mu$m$^2$. 

\begin{figure*}[t]
\def\ffile{Reproduce.pdf}
\centering
\includegraphics[width=0.8\linewidth]{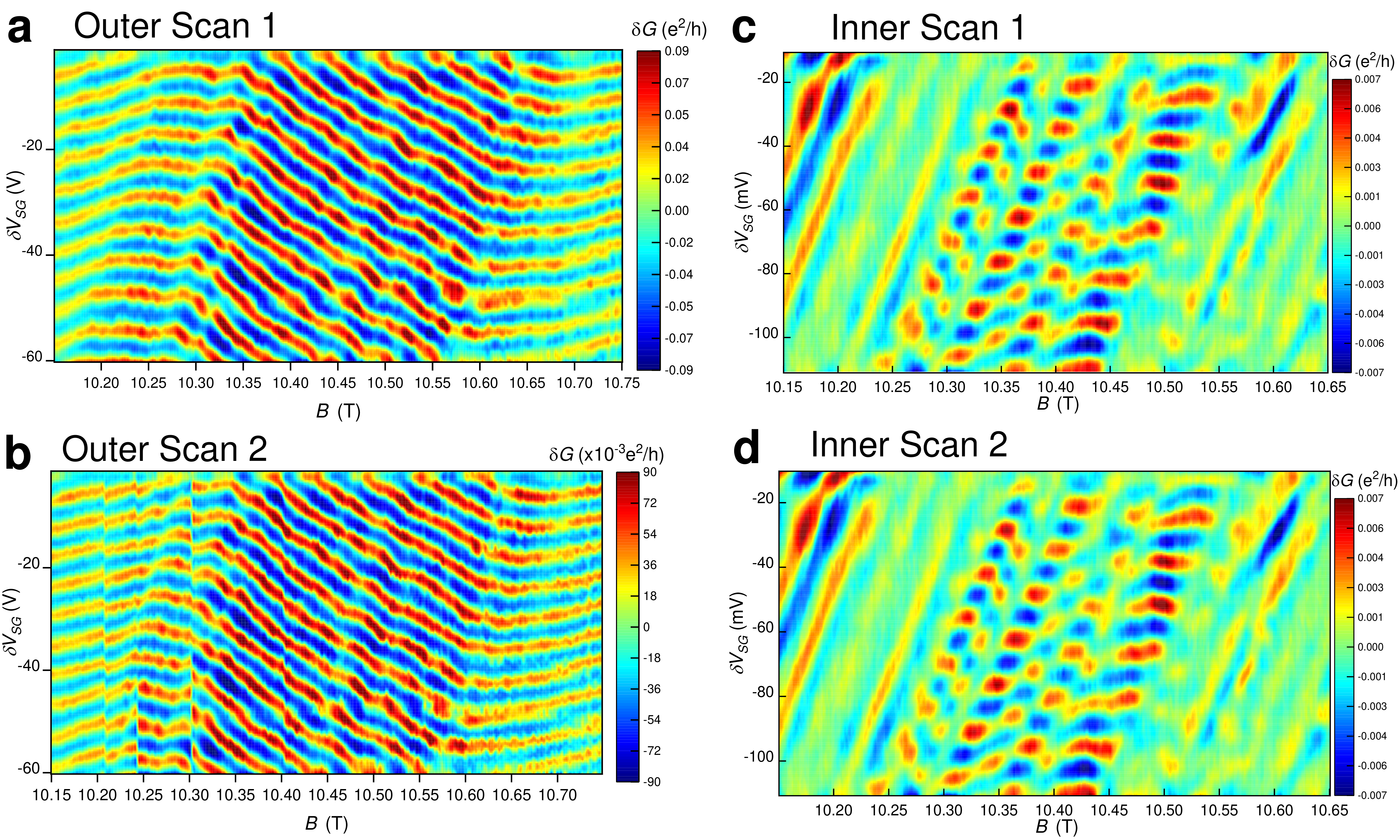}
\caption{\label{Reproduce}Repeatability of interference in Device A. a) Scan 1 for the outer mode at $\nu = 2/5$, repeated from the main text Fig. 3. b) Second scan for the outer mode, using the same gate voltages and no changes in device operating parameters. Scan 1 is taken starting from low magnetic field and moving to high field, whereas scan 2 starts at high field and moves to low field. Scan 2 happens to have a few charge switching events, likely caused by tunneling of charge in the doping layer. c) Scan 1 for inner mode, repeated from main text Fig. 5. b) Scan 2 using the same scan paramaters as c) for the inner mode. Scan 1 steps the field from low field to high field, while scan 2 steps the field from high field to low field.}
\end{figure*}

The side gate will also couple slightly to charge in the central region of the interferometer, which we parameterize through $\alpha _{bulk} = \frac{\partial \Bar{q_b}}{\partial V_{SG}}$. In this case $\Bar{q_b}$ is the effective background charge, and any increase in $\Bar{q_b}$ will lead to additional quasiparticles if the bulk is {\it compressible}, and may lead to changes in the interferometer area via bulk-edge coupling. To extract $\alpha_{bulk}$, we note that the total lever arm relating change in charge on the dot to change in gate voltage is $\alpha_{total} = \frac{\partial q_{total}}{\partial V_{SG}} = \alpha_{edge}+\alpha_{bulk}$. Note $q_{total}=q_b + q_{edge}$. When the device is at zero magnetic field and there is no gap, there is no meaningful distinction between charge in the bulk and charge on the edge; thus, Coulomb blockade oscillations, which have conductance peaks corresponding to removal of single electrons from the dot, have a period $\Delta V_{SG} = \frac{1}{\alpha _{total}}$. We have measured Coulomb blockade through the interferometer at $B = 0$ with the device tuned to the weak tunneling regime. In this configuration $\Delta V_{SG} = 5.6$ mV, yielding $\alpha _{total} = 0.179$ mV$^{-1}$. $\alpha _{edge}$ can be calculated from the product of electron density and change in area with gate voltage, $\alpha _{edge} = \frac{\partial \Bar{A}}{\partial V_{SG}} \times n = 0.12$ mV$^{-1}$. Then, $\alpha _{bulk} = \alpha _{total} -\alpha_{edge} = 0.179$ mV$^{-1}$ $-0.12$ mV$^{-1} = 0.06$ mV$^{-1}$.

\section{Supplementary Section 3: Reproducibility of Interference Signal}

We have repeated each of the scans of interference of the inner and outer modes in Device A. Fig. \ref{Reproduce}a and b show two consecutive scans of interference for the outer mode (Fig. \ref{Reproduce}a is reproduced from main text Fig. 3). The gate voltages and scan parameters are the same; the only difference is the scan direction, with Scan 1 going from low field to high field and Scan 2 from high field to low field. The two scans show nearly the same behavior, except that there are a few charge switching events in Scan 2. These are likely caused by charge noise in the doping layer, a known issue in doped GaAs/AlGaAs devices. Since interference of the outer mode requires more negative QPC gate voltages, such charge switching events are somewhat more common.

Fig. \ref{Reproduce}c and d show two consecutive scans for the inner mode at $\nu = 2/5$. Scan 1 is taken increasing the field from low field to high field, while Scan 2 is taken decreasing the field from high field to low field, with the other parameters being the same. The behavior is nearly the same in both scans, indicating that the series of discrete jumps in phase in phase are repeatable features.

\section{Supplementary Section 4: Extracting phase via Fourier transform}
\begin{figure*}[t]
\def\ffile{Phase_Extraction.pdf}
\centering
\includegraphics[width=1.0\linewidth]{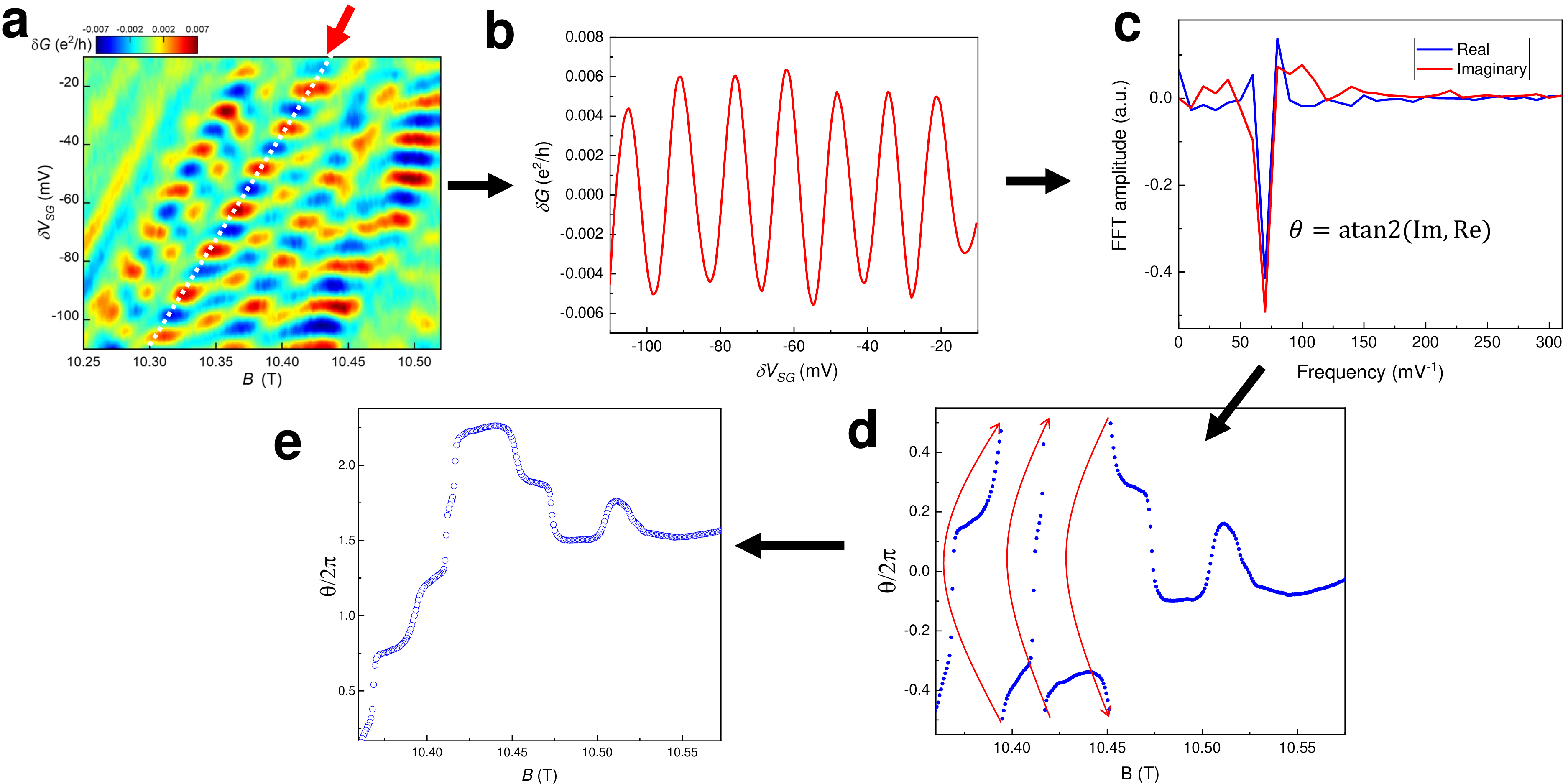}
\caption{\label{Phase_Extraction}a) Conductance variation versus magnetic field $B$ and side gate voltage $\delta V_{SG}$ for the inner mode at $\nu = 2/5$, reproduced from the main text. b) Line cut along a diagonal parallel to the discrete jumps (indicated with the red arrow and white dashed line in a). c) Real and imaginary parts of a Fast Fourier Transform of the data in b). There is a clear peak in both components corresponding to the interference oscillation period, and the phase is extracted from the inverse tangent of the real and imaginary amplitudes. d) Phase extracted at each value of $B$. Since the phase from inverse tangent is from $-\pi$ to $+\pi$, all the values fall in this range. e) In order to avoid discontinuities, the at each shift from $-0.5\times2\pi$ to $+0.5\times 2\pi$ the phase is shifted up, as illustrated with red arrows in d).}
\end{figure*}
In order to extract the interference phase from conductance data, we take Fourier transforms across diagonal cuts of conductance (this is the same process used to extract the phase in Ref. \cite{Nakamura2022}. The process is illustrated in Fig. \ref{Phase_Extraction}. Line cuts of conductance along diagonals parallel to the discrete jumps are taken; this way the phase in regions of constant quasiparticle number can be extracted from a Fourier transform. The phase from the Fourier transform is defined from $-\pi$ to $+\pi$; to avoid discontinuities around $\pm \pi$ we shift the values of phase as illustrated in Fig. \ref{Phase_Extraction}d, yielding the curve of Fig. \ref{Phase_Extraction}e (the same as is shown in the main text for the inner mode at $\nu = 2/5$. 

In general the resulting values of the phases should include both the discrete jumps in phase due to a change in quasiparticle number as well as the continuously varying Aharonov-Bohm phase, which increases linearly with $B$ and will give a linear slope. This linear slope will be positive in the AB regime and negative in the CD regime. However, for the inner mode at $\nu = 2/5$, the lines of constant phase are nearly flat in the regions between the discrete jumps, as discussed in the main text, which is due to the bulk-edge coupling placing the device right at the edge of the AB-CD transition. Therefore, for the analysis of $\nu = 2/5$, we have not subtracted the Aharonov-Bohm slope. This lack of $B$ dependence to the oscillations can be seen from the fact that in Supp. Fig. \ref{Phase_Extraction}e, the plateaus in phase are nearly flat even without any AB slope subtraction. In some regions (particularly at lower field) there may be a slight slope, but this very weak $B$ dependence will not affect the values of the discrete jumps extracted very strongly, so for simplicity in our analysis we have not subtracted any slope. For the analysis of $\nu = 1/3$ in later parts of the supplementary material, on the other hand, we extract the phase in the same way, but we do subtract the continous AB slope since there is clear negatively-sloped AB behavior.

\section{Supplementary Section 5: Simulations}
\begin{figure*}[t]
\def\ffile{Simulations_flat.pdf}
\centering
\includegraphics[width=1.0\linewidth]{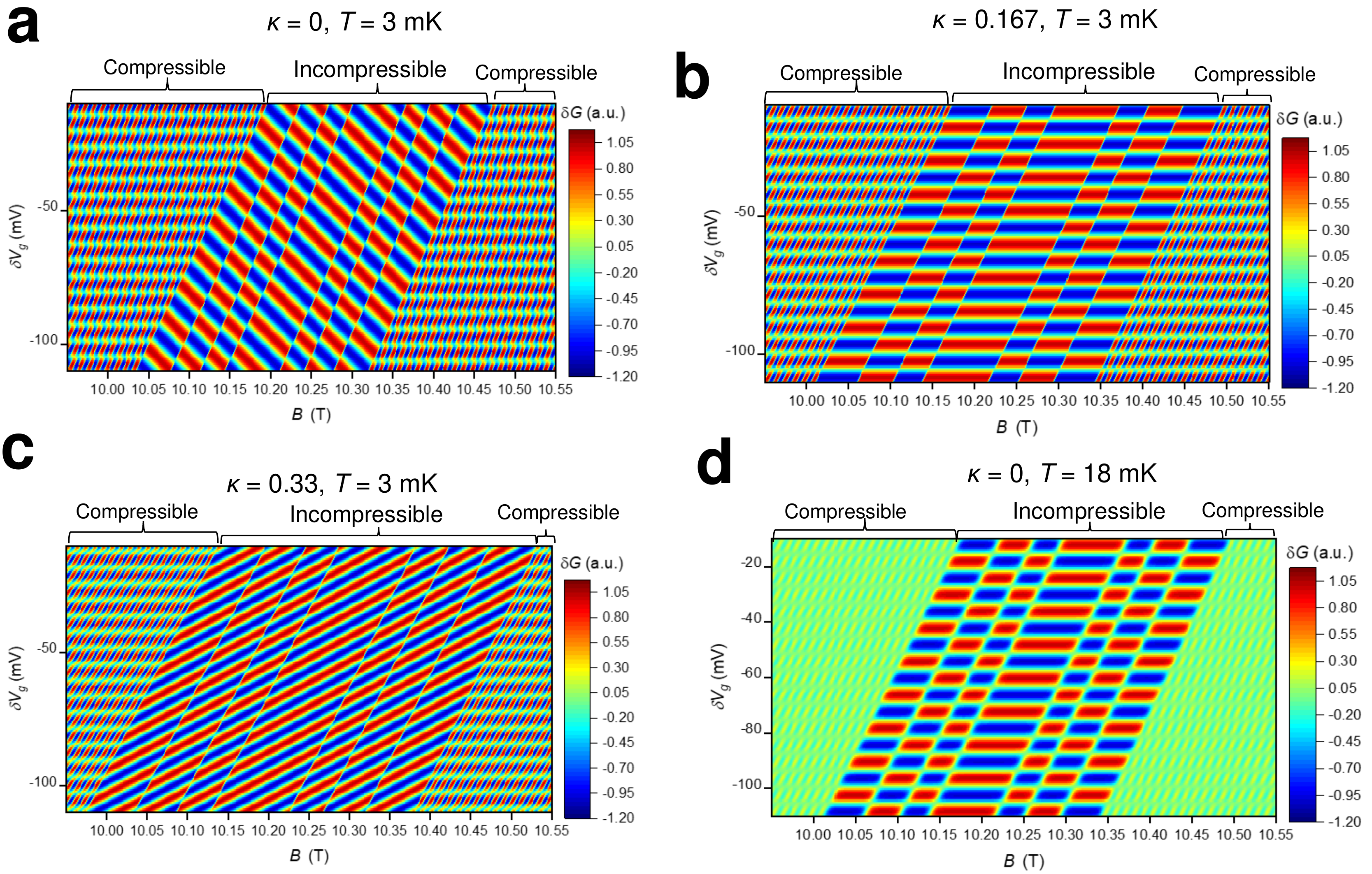}
\caption{\label{Simulations}Simulations of $\nu = 2/5$ interference. a) Simulation of conductance for the inner mode with zero bulk edge coupling at low temperature $T = 3$ mK. Lines of constant phase have negative slope in the incompressible regime, and in the compressible regime the discrete jumps with $\frac{\Phi_0}{2}$ magnetic field period due to anyonic statistics create a checkerboard pattern with nearly vertical average lines of constant phase. b) Simulation with finite bulk edge coupling $\kappa = 0.167$, right at the transition from the AB to the Coulomb-dominated regime. In the incompressible region the lines of constant phase have no magnetic field dependence, similar to the experimentally observed behavior in Main Text Fig. 5a. c) Simulation with $\kappa = 0.33$, in the Coulomb dominated regime. In the incompressible region the lines of constant phase have positive slope. d) Simulation with $\kappa = 0.167$ and $T = 18$ mK. In the compressible regions, the quasiparticle number is strongly thermally smeared, resulting in a dramatic reduction in interference amplitude.}
\end{figure*}

Following Ref. \cite{Halperin2011} and Main Text Eqn. 2, we have performed simulations of conductance oscillations for the inner mode as a function of gate voltage and magnetic field for our device, as shown in Fig. \ref{Simulations}. These simulations assume an average area $\bar{A} = 0.4$ $\mu$m$^{2}$, an electrostatic charging energy 50 $\mu e$V set by the setback of the screening layers (close to what we have measured in $B=0$ Coulomb blockade, $e^* = 1/5$, and $\theta _a = -\frac{4\pi}{5}$; these parameters approximately match those of Device A. Different values for $\kappa$ have been simulated, with $\kappa = 0$ (pure AB) shown in Fig. \ref{Simulations}a, $\kappa = 0.167$ (at the transition from AB to CD) in Fig. \ref{Simulations}b, and $\kappa = 0.33$ (CD regime) in Fig. \ref{Simulations}c. An energy gap of 2 K has been assumed in order to match the approximate width of the incompressible region (somewhat smaller than the activation gap measured in bulk transport of 3.5 K). The simulations include thermal smearing of the quasiparticle number; Fig. \ref{Simulations}a, b, and c are at a low temperature of $T = 3$ mK which results in minimal thermal smearing, while Fig. \ref{Simulations}d is at $\kappa = 0.167$ and $T = 18$ mK, close to the electron temperature we have inferred from $B = 0$ Coulomb blockade (this is somewhat higher than the mixing chamber temperature of 10 mK, which is typical for measurements of mesoscopic devices). The simulations do not include thermal smearing of the Aharonov-Bohm phase with temperature, though this will also lead to an overall reduction in amplitude at higher $T$ (but should not qualitatively change the interfernece pattern). In order to illustrate the effect of disorder, quasiparticle states with energies of -0.7, -0.45, -0.25, +0.25, +0.45, and +0.7 (in units of the gap) have been included, which result in a few isolated discrete phase jumps in phase, similar to the experimentally observed behavior.

As discussed previously, in the experimental data of Main Text Fig. 5a, the lines of constant phase in the incompressible region in the regions between the discrete jumps are nearly flat as a function of magnetic field. This is unlike the simulations at $\kappa = 0$ or $\kappa = 0.33$, and most closely resembles the simulation at $\kappa = 0.167$. The absence of $\Phi_0$ period compressible regime interference in the experiment is consistent with the $T=18$ mK simulation in Supp. Fig. \ref{Simulations}d in which oscillations in the compressible region have an amplitude that is suppressed by approximately a factor of 10 due to thermal smearing of the quasiparticle number. Since the experimentally observed amplitude is already very low in the {\it incompressible} region, it is reasonable that interference may be too small to measure when the bulk becomes compressible, and only resonances due to conduction through the bulk are visible.

\section{Supplementary Section 6: Analysis of bulk-edge coupling via finite-bias measurements}

\begin{figure*}[t]
\def\ffile{Finite_bias_flat.pdf}
\centering
\includegraphics[width=1.0\linewidth]{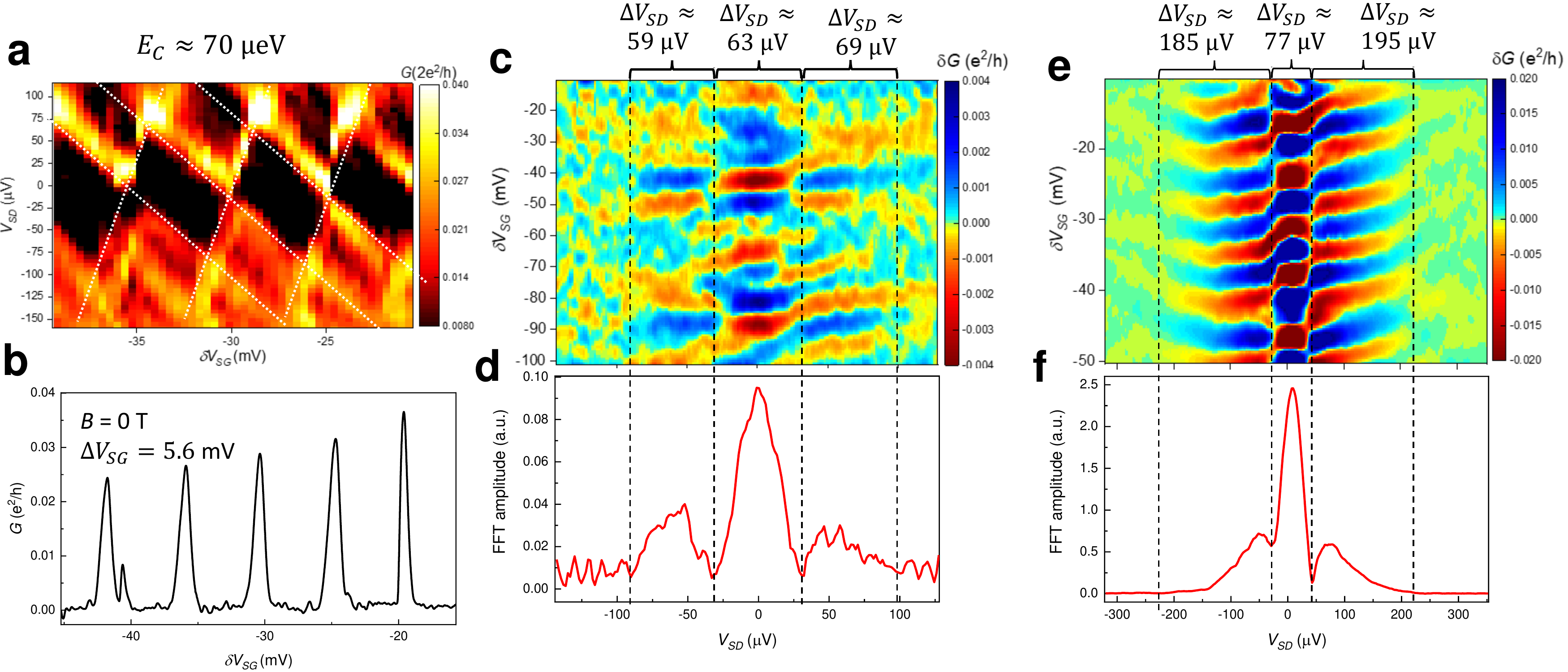}
\caption{\label{Fin}Finite bias measurements of Device A. a) Coulomb diamond plot of conductance versus side gate voltage and $V_{SD}$ at $B = 0$. The height of the Coulomb blockade diamonds gives $\Delta E_{CD} \approx 70$ $\mu$eV. b) Coulomb blockade oscillations with gate voltage period $\Delta V_{SG} = 5.6$ mV. c) Differential conductance as a function of $\delta V_{SG}$ and $V_{SD}$ with QPCs tuned to probe the inner mode at $\nu = 2/5$. d) Oscillation amplitude (extracted via FFT of $\delta G$ vs $\delta V_{SG}$) as a function of source-drain bias $V_{SD}$. The spacing of the outer nodes in the oscillation amplitude give a voltage scale $\Delta V_{SD}\approx 64$ $\mu e$V. e) Differential conductance for outer mode. f) FFT amplitude versus $V_{SD}$ for outer mode. The spacing of the outer nodes in the oscillation amplitude give a voltage scale $\Delta V_{SD}\approx 190$ $\mu e$V. }
\end{figure*}

Due to the significant impact of bulk-edge coupling on phase at $\nu = 2/5$, extracting the values of $e^*$ and $\theta_a$ requires knowledge of the parameter $\kappa$. Evaluation of $\kappa$ is somewhat more complicated at $\nu=2/5$ compared to $\nu = 1$ and $\nu = 1/3$ which have single edge states because there may be an interaction between the two edge states that impacts $\kappa$. Finite bias measurements can be used to extract relevant coupling parameters \cite{Nakamura2022}; here we extend this method to the two edge state case of $\nu = 2/5$. Other approaches based on oscillation periods have been used to estimate bulk-edge coupling parameters \cite{Sivan2016a, Roosli2020, Roosli2021}, but applied only for oscillations of the outer edge states. Theoretical analysis of interferometers with two integer edge states has been undertaken \cite{Frigeri2019}, but this analysis also focused on the case of interfering only the outer edge mode.

In the case of a single edge, the electrostatic energy associated with variations in charge on the edge and bulk has been approximated in the following form \cite{Halperin2011, VonKeyserlingk2015}:

\begin{equation}
\label{E_single}
E = \frac{k_e}{2}(\delta q_e)^2 + \frac{k_b}{2}(\delta q_b)^2+k_{be}\delta q_b \delta q_e
\end{equation}

Where $\delta q_e\equiv \frac{\delta A \nu B}{\Phi_0}$ and $\delta q_b \equiv \frac{\Bar{A} \nu B}{\Phi_0}+e^* N_{qp}-\Bar{q}$ are the variations in charge on the edge and the bulk respectively (in units of the electron charge $e$). $k_e$ is the stiffness of the edge that describes the energy cost of changing the amount of charge on the edge, and $k_b$ is the charging energy of the bulk, and $k_{be}$ is the bulk-edge coupling constant. In this case, minimizing the energy in response to a change in bulk charge $\delta q_b$ (which could occur due to a change in condensate charge $\frac{\Bar{A} \nu B}{\Phi_0}$ or from changes in localized quasiparticle number $N_{qp}$) gives a change in edge charge $\kappa = -\frac{\delta q_{e}}{\delta q_b}= \frac{k_{be}}{k_e}$.

The case of two edges is more complicated because in principle the two edges should interact, making each edge partially screen the other so that each edge will have a smaller change in charge in response to a perturbation of the bulk. In principle, a model should include coupling constants between the edges as well as between each edge and the bulk. As an approximation, we model the device as having a single electrostatic interaction energy $E_{int}=k_{int} (\delta q _{total})^2$ similar to the approach of Ref. \cite{Rosenow2007}. In this approximation, the effect of the interaction energy is to limit the total buildup of excess charge in the device. Since the edge charge variation will be distributed approximately uniformly across the perimeter of the device, the typical interaction energy between charge a charge on one edge with a charge on the other will not be very different from the interaction of the edge with charge in the bulk unless the edges are very close together. The fact that there is a reasonably wide intermediate plateau at $G = \frac{1}{3} \frac{e^2}{h}$ in the individual QPC sweeps in Main Text Fig. 2b indicates that there is an appreciable separation between the two edge modes, making this a reasonable approximation. Additionally, each edge will have a ``single particle'' contribution to the energy ($E_{sp1}$ and $E_{sp2}$ for the inner edge and outer edge), which reflects the energy cost to add charge to the edge due to the external confining potential (in other words, this contribution to the energy represents the velocity without interaction with the other modes, although the actual propagation velocities will be modified by the interaction). In this approximation, the energy with two modes is given by:

\begin{equation}
\label{E_two}
\begin{split}
E = E_{1sp}+E_{2sp}+E_{int} \\ = \frac{k_{1sp}}{2}(\delta q_1)^2 +\frac{k_{2sp}}{2}(\delta q_2)^2+ \frac{k_{int}}{2}(\delta q_{total})^2
\end{split}
\end{equation}

In Eqn. \ref{E_two}, $\delta q_1 \equiv \frac{\delta A_1 \Delta \nu_1 B}{\Phi_0}$ is the variation in charge on the inner edge (in units of electron charge $e$), $\delta q_2 \equiv \frac{\delta A_2 \Delta \nu_2 B}{\Phi_0}$ is the variation on the outer edge, and $\delta q_{total}\equiv \delta q_1+\delta q_2+\delta q_b$ is the total variation in charge on the device. For the case of $\nu = 2/5$, $\delta A_1$ is the variation in area of the region enclosed by the inner edge state, $\Delta \nu _1 = \frac{2}{5} -\frac{1}{3}=\frac{1}{15}$ is the effective filling factor of the inner condensate, $\delta A_2$ is the variation in area enclosed by the outer edge state, and $\Delta \nu _2 = \frac{1}{3} -0=\frac{1}{3}$ is the effective filling factor of the outer condensate (i.e. the outer incompressible strip that separates the inner and outer edge states).

The interaction contribution to the energy $E_{int}=k_{int} (\delta q_{total})^2$ can be extracted from the Coulomb blockade diamond pattern \cite{Beenakker1991}, shown in Fig. \ref{Fin}a. The height of each diamond is the energy required to add an additional electron to the dot, which is a combination of the zero-field level spacing $\Delta E_{sp, B=0}$ and the electrostatic charging energy, which is just the interaction energy $k_{int} (\delta q_{total})^2$. So, the height of each Coulomb diamond in terms of energy is $\Delta E_{CD}=\Delta E_{sp, B=0}+k_{int}$. Since the device is large compared to the Fermi wavelength, we approximate the $B=0$ single particle level spacing from the density of states in 2D \cite{Beenakker1991, Kouwenhoven1999}, $\Delta E_{sp, B=0}= \frac{\pi \hbar ^2 A }{m^*}\approx 10$ $\mu e$V. Since $\Delta E_{CD} = 70$ $\mu e$V from Supp. Fig. \ref{Fin}a, $k_{int} = 70$ $\mu e$V - 10 $\mu e$V = 60 $\mu e$V.

Chamon et al. \cite{Chamon1997} showed that, in quantum Hall interferometers with a single edge mode, conductance oscillates as a function of source-drain bias $V_{SD}$ because phase depends on energy, with a change in quasiparticle wavevector $\delta k= \nu \frac{\delta E}{\hbar v_{edge}}$ for a given change in energy $\delta E$ for an edge state with velocity $v_{edge}$, and thus a shift in phase $\Delta \theta = \frac{L \delta E}{\hbar v_{edge}}$ (with $L$ the perimeter of the interference path). In the case of a symmetric potential drop, this results in nodes in the conductance at certain values of $V_{SD}$; for fractional quantum Hall edge states the innermost pair of nodes have are expected to have a narrower spacing which is sensitive to temperature due to the Luttinger liquid nature of the edge \cite{Wen1991a, Chamon1997}, while the outer nodes have approximate spacing $\Delta V_{SD} = \frac{h v_{edge}}{ee^*L}$ (this narrower spacing of the central nodes with approximately uniform outer node spacing has been observed at $\nu = \frac{1}{3}$ \cite{Nakamura2022}). Since the velocity is determined by the electric field $v_{edge}= \frac{\mathcal{E}\times B}{B^2}$, this edge velocity implies an electric field magnitude $\mathcal{E}=B v_{edge}$, and thus the edge stiffness $k_{e}=\frac{e\Phi_0 \mathcal{E}}{LB \Delta \nu}$ (the factor in the denominator of $L B\Delta \nu$ accounts for how much the edge must expand against the electric field $\mathcal{E}$ in order to add one unit of charge to the edge).

For a state with multiple edges that interact, however, this situation will be complicated because interactions between the edge states will result in mixed eigenmodes with different velocities. A discussed in Ref. \cite{Smits2013}, in this case multiple frequencies may occur in the conductance as a function of $V_{SD}$ with frequencies $\frac{ee^*L}{2v_{i}h}$, with $v_i$ the velocity of each (mixed) eigenmode (even more frequencies will occur if the path length on each edge is different, but because our device is symmetric, this effect should not be relevant). This implies that the spacing of nodes corresponding to each frequency (half the oscillation period) will be $\Delta V_{SD} = \frac{h v_{edge}}{ee^*L}$, the same as for the single edge case in Ref. \cite{Chamon1997}, but with $v_i$ the velocity of the eigenmodes. 

The matrix equation for density variations along the edge \cite{Fujisawa2022}: 

\begin{equation}
\label{Modes}
v_i  \begin{pmatrix}
\rho_1 \\ \rho_2
\end{pmatrix}  = 
\begin{pmatrix}
G_1 & 0\\
0 & G_2
\end{pmatrix}
\begin{pmatrix}
K_{1sp} & K_{12} \\
K_{12} & K_{2sp}
\end{pmatrix}
\begin{pmatrix}
\rho_1 \\
\rho_2
\end{pmatrix}
\end{equation}

With $G_1 = \frac{1}{15}\frac{e^2}{h}$ and $G_2= \frac{1}{3}\frac{e^2}{h}$ the conductance of the inner and outer edge states, $K_{1sp}$ and $K_{2sp}$ the individual edge stiffness of each edge state, and $K_{12}$ the interaction between the edge states normalized for length. In the approximation of Eqn. \ref{E_two}, $K_{1sp} = \frac{k_{1sp}L}{e^2}$, $K_{2sp} = \frac{k_{2sp}L}{e^2}$, and $K_{12} = \frac{k_{int}L}{e^2}$ (the factor of the perimeter $L$ normalizes the parameters to the length of the edge of the device). Solving this eigenvalue equation gives the velocities of the propagating eigenmodes.

The fact that the spacings of the nodes for the inner and outer edge states are very different suggests that each edge state corresponds primarily to one of the two propagating modes, i.e. the interactions do not cause strong mixing. The factor of $G_1$ and $G_2$ in Eqn. \ref{Modes} will tend to limit mixing because the two edge states have very different conductances. Therefore, it is reasonable to expect that the modification of the velocities due to interactions between the modes will not be large. The eigenvalues of Eqn. \ref{Modes} are:

\small
\begin{equation}
\label{mode_velocity}
v_i = \frac{G_1 K_{1sp}+G_2 K_{2sp} \pm \sqrt{(G_1 K_{1sp} -G_2 K_{2sp})^2 +4G_1 G_2 K_{12}^2}}{2}
\end{equation}
\normalsize
The oscillation amplitude for both the inner and outer mode decreases dramatically as $V_{SD}$ increases, as can be seen from Fig. \ref{Fin}c and e. This introduces some uncertainty in the positions of the nodes; nevertheless, node-like features do appear, and the spacing of the outer nodes can be used to estimate the eigenmode velocities. The velocity extracted from the outer node spacing when interfering the inner mode in Fig. \ref{Fin}c is $v_1 = \frac{ee^* \Delta V_{SD}L}{2h} \approx \frac{e}{5} \frac{64 \mathrm{\mu V} \times 2.5 \mathrm{\mu m}}{h} \approx 7.7 \times 10^3$ m/s (assuming tunneling charge $e/5$ expected for the inner mode), while for the outer mode in Fig. \ref{Fin}e it is \ref{Fin}c is $v_2 = \frac{ee^* \Delta V_{SD}L}{2h} \approx \frac{e}{3} \frac{190 \mathrm{\mu V} \times 2.5 \mathrm{\mu m}}{h} \approx 3.8 \times 10^4$ m/s (assuming tunneling charge $e/3$ for the outer mode). The significantly higher velocity for the outer mode is consistent with the expectation of steeper confining potential at the outer edge of the 2DEG \cite{Chklovskii1993, Sahasrabudhe2018, Nakamura2019}. Using the value of $k_{int} \approx 60$ $\mu$eV extracted from Coulomb blockade to approximate the interaction yields $K_{12}=9.4\times 10^8$ Vm/C. The values of $K_{1sp}$ and $K_{2sp}$ are straightforward to calculate from Eqn. \ref{mode_velocity}, yielding $G_1 K_{1sp} = 9\times 10^3$ m/s and $G_2 K_{2sp} = 3.7\times 10^4$ m/s. Then, $k_{1sp} = 223$ $\mu e$V and $k_{2sp} = 183$ $\mu e$V. Note that these values are close to the values of $\frac{2h v_1}{e^* L}=\Delta V_{SD, 1}$ and $\frac{2h v_2}{e^* L}$  from the differential conductance measurements, indicating that there is only weak mixing of the modes. This justifies extracting only a single frequency (i.e. a single value of $\Delta V_{SD}$) from the differential conductance measurements, whereas with strong mixing the velocities would be significantly different and multiple frequencies would contribute significantly. Additional theoretical investigation could give more insight into the mixing of fractional edge states by interaction.

These values can be applied to Eqn. \ref{E_two} to calculate $\kappa$. Minimizing the energy of Eqn. \ref{E_two} with respect to $\delta q_1$ and $\delta q_2$ for a given $\delta q_b$ by setting $\frac{\partial E}{\partial \delta q_1} = 0$ and $\frac{\partial E}{\partial \delta q_2} = 0$ and solving for $\delta q_1$  yields 

\begin{equation}
    \label{kappa}
    \kappa \equiv -\frac{\delta q_1}{\delta q_b} = \frac{k_{2sp}k_{int}}{k_{1sp}k_{2sp}+k_{int}k_{1sp}+k_{int}{k_2{sp}}}
\end{equation}

Applying the extracted interaction parameters to Eqn. \ref{kappa} yields $\kappa \approx 0.17$, very close to the critical value of $\frac{1}{6}$ at which the lines of constant phase in the compressible region become flat in our simulations.

Using Main Text Eqn. 4, this value of $\kappa$ can be used to extract the anyonic phase from the value of the phase jumps by subtracting the contribution from bulk-edge coupling, giving $\theta _a = -\Bar{\Delta \theta}+\frac{3}{5}\kappa \approx -0.43 \times 2\pi$, close to the theoretically predicted value of $\theta_a = -\frac{2}{5}\times 2\pi$. It is noteworthy that there is some variation in phase in the discrete jumps in Main Text Fig. 6, and all of them except for one fall into the range described by Main Text Eqn. 4 with values between $\frac{2}{5}\times 2\pi$ and $2\pi$. The one outlier is jump number 7, which has a value of $0.25\times 2\pi$. There are a few possible reasons why this jump falls outside this range; this could be due to inaccuracies in the extracted phase (it appears that at ~10.47 T the interference amplitude decreases, which might cause some additional inaccuracy compared to other jumps). Qualitatively, it also appears that jumps 7 and 8 smear together somewhat, which makes the region in between the two jumps less well defined, which could make the extracted value of the jump smaller than the true value. Another interesting possibility is that jump 7 might correspond not to an $e/5$ quasiparticle in the inner condensate, but an $e/3$ quasiparticle localized in the incompressible $\nu = 1/3$ region. Assuming the quasiparticles in that region have approximately the same coupling to the inner edge as the localized charge in the inner condensate does, a phase jumps of $\Delta \theta = \kappa \frac{e^*_i e^*_{local}}{\Delta \nu} = 15\kappa \frac{1}{5}\frac{1}{3}=\kappa$, with $e^*_i = \frac{1}{5}$ the interfering charge and $e^*_{local}=\frac{1}{3}$ being the localized charge in this situation. This would yield a phase jump $\Delta \theta = 0.17\times 2\pi$, fairly close to the observed value of 0.25.

It is worth emphasizing that the extraction of edge velocities and interaction parameters here relies on a simplified model with several approximations. Future theoretical investigation could lead to better understanding of the connection between finite-bias measurements, edge velocities, and Coulomb coupling parameters in interferometers, leading to more precise determinations of $\kappa$.

\section{Supplementary Section 7: Dependence of $\kappa$ on inter-edge coupling}
In the previous section, we have approximated the coupling of the two edges to each other as being equal to the bulk-edge coupling, meaning that there is a single parameter $k_{int}$ that describes all charge interactions in the device. However, this likely somewhat underestimates the interaction of the two edges, since they are located close together. A higher value of the edge-edge interaction will affect the edge stiffness parameters extracted from Supp. Eqn. \ref{mode_velocity} as well as the degree of screening of the inner edge by the outer edge, and will thus impact the calculated value of $\kappa$.

To determine the sensitivity of our analysis to the value of the inter-edge coupling, we have calculated $\kappa$ assuming a large edge-edge coupling of 120 $\mu e$V, double the value of the electrostatic charging energy. This results in $K_{12}=1.88\times 10^9$ Vm/C. From Supp. Eqn. \ref{mode_velocity}, $G_1 K_{1sp} = 1.2\times 10^4$ m/s and $G_2 K_{2sp} = 3.6\times 10^4$ m/s. The single-particle edge stiffnesses are $k_{1sp} = 303$ $\mu e$V and $k_{2sp} = 167$ $\mu e$V. Minimizing the energy with these parameters (including the larger edge-edge interaction $k_{12}$) yields $\kappa \approx 0.08$. 

Using this value of $\kappa$ that assumes larger edge-edge interaction yields $\theta _a = -\Bar{\Delta \theta}+\frac{3}{5}\kappa \approx -0.48 \times 2\pi$, somewhat larger than the value of $\theta _a = -0.43\times 2\pi$ extracted when the inter-edge interaction was assumed to be equal to the bulk-edge interaction, but the difference is not dramatic. Since a factor of two difference in the edge-edge interaction only results in a modest change in the extracted $\theta_a$, our analysis is likely not overly sensitive to errors from the simplification we made by treating all interactions as equal in Supp. Section 6. It is also noteworthy that the fact that the phase in the regions in between the discrete jumps are nearly flat and independent of $B$ is more consistent with the value of $\kappa =0.17$ obtained from Supp. Section 6, suggesting that the approximations made are reasonable. Nevertheless, as discussed in the previous section, additional theoretical investigation of bulk-edge coupling when multiple interacting edge modes are present is warranted, and could improve the estimate of $\kappa$.


\section{Supplementary Section 8: Temperature dependence at 2/5}
\begin{figure*}[t]

\def\ffile{Temp_flat.pdf}
\centering
\includegraphics[width=0.7\linewidth]{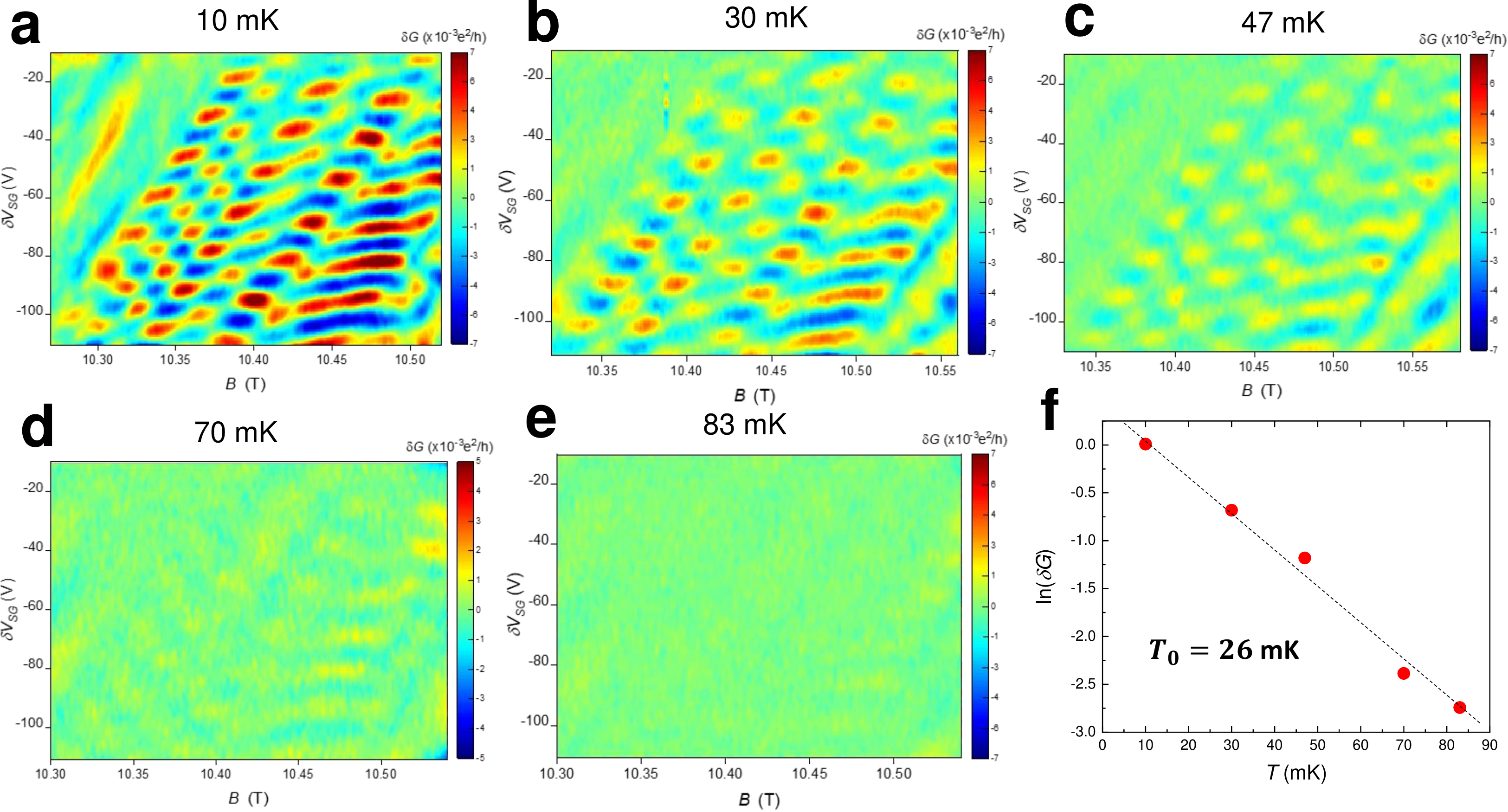}
\caption{\label{Temp}Interference of the inner mode at different cryostat temperatures. a) 10 mK b) 30 mK c) 47 mK d) 70 mK e) 83 mK. f) Natural log of the amplitude versus temperature. A linear fit yields a temperature scale $T_0=26$ mK}
\end{figure*}

Fig. \ref{Temp} shows measurements of the inner mode at different cryostat temperatures in Device A. While the behavior does not change qualitatively, the oscillation amplitude drops quite rapidly with increasing temperature, becoming nearly unmeasurable past $T=83$ mK. Fig. \ref{Temp}f shows the natural log of the amplitude (relative to the amplitude at base temperature $T = 10$ mK) versus temperature; the data is approximately linear, indicating that the amplitude decays exponentially with temperature. A linear fit yields a temperature scale $T_0 = 26$ mK. This is much smaller than the temperature scale of $T_0=94$ mK measured in a similarly-sized device at $\nu = 1/3$ in the incompressible regime \cite{Nakamura2020}, indicating that interference at $\nu = 2/5$ is significantly more fragile. This is probably due to the low edge velocity of the inner mode (as discussed in the previous section). 

\section{Supplementary Section 9: Interference at $\nu = 1/3$}
\begin{figure}[t]
\def\ffile{nu_1_3_jumps_flat.pdf}
\centering
\includegraphics[width=1.0\linewidth]{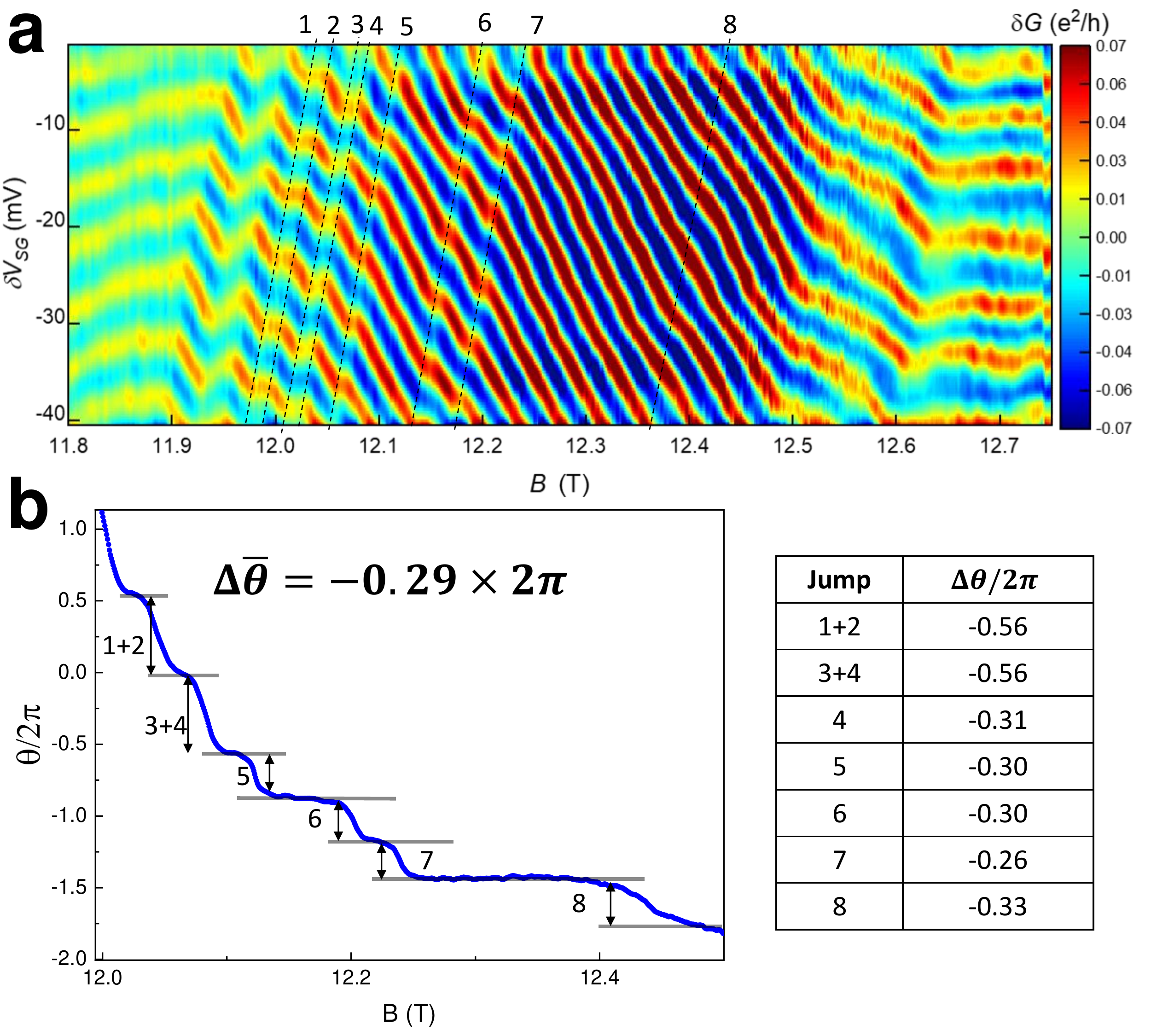}
\caption{\label{nu_1_3}Interference at $\nu = 1/3$ in Device A. a) Conductance versus $B$ and $\Delta V_{SG}$ at $\nu = 1/3$. There is a wide central incompressible region where the device exhibits negatively sloped AB oscillations with a few discrete phase jumps, and at high and low field the bulk becomes compressible and the lines of constant phase become nearly flat. b) Phase versus magnetic field extracted via Fourier transform. A constant slope from the Aharonov-Bohm contribution to the phase is subtracted in order to isolate the contribution from the discrete jumps. The values of the phase jumps are listed in the table.}
\end{figure}

Fig. \ref{nu_1_3}a shows interference at $\nu = 1/3$ for Device A. This is a different measurement than main text Fig. 4; whereas the measurement shown in the main text involved smoothly adjusting the QPC voltages so that continuous interference from 2/5 to 1/3 could be achieved, here the QPC voltagese are kept fixed since the measurement is only at 1/3. The behavior is consistent with what has been previously reported in screening well interferometer \cite{Nakamura2020, Nakamura2022}: there is a central incompressible region where there are negatively sloped lines of constant phase and a few discrete jumps due to anyonic statistics, and at high and low field the lines of constant phase flatten out as the bulk becomes incompressible and many quasiparticles/quasiholes are created. Discrete jumps in phase are indicated with dashes lines and numbered.

The phase extracted via FFT are shown in Fig. \ref{nu_1_3}b. The continuously-varying Aharonov-Bohm contribution has been subtracted off. The discrete jumps in phase show up as negative steps in phase. It is notable that jumps 1 and 2 are very close together and form nearly one continuous step in phase, and jumps 3 and 4 also do this. These can be distinguished from single jumps by the wider extent of the transition between the flat regions, and by the fact that change in phase is still negative (whereas for a single jump with the total change in phase for jumps 1+2 or 3+4, the jump would appear to be a positive change in phase since it exceeds -0.5). The average value of the phase jumps is $-0.29 \times 2\pi$, close to the value of $-0.31\times2\pi$ previously reported for a device of similar dimensions \cite{Nakamura2020}, and close to the value of $-\frac{2\pi}{3} = -\theta_a$ expected for removal of an $e/3$ anyonic quasiparticle. Note that the magnitude is slightly smaller than the ideal value, consistent with the expected effect of finite bulk-edge coupling at $\nu = 1/3$.

\section{Supplementary Section 10: Device B}

\begin{figure*}[t]
\def\ffile{Inner_800nm_v2.pdf}
\centering
\includegraphics[width=1.0\linewidth]{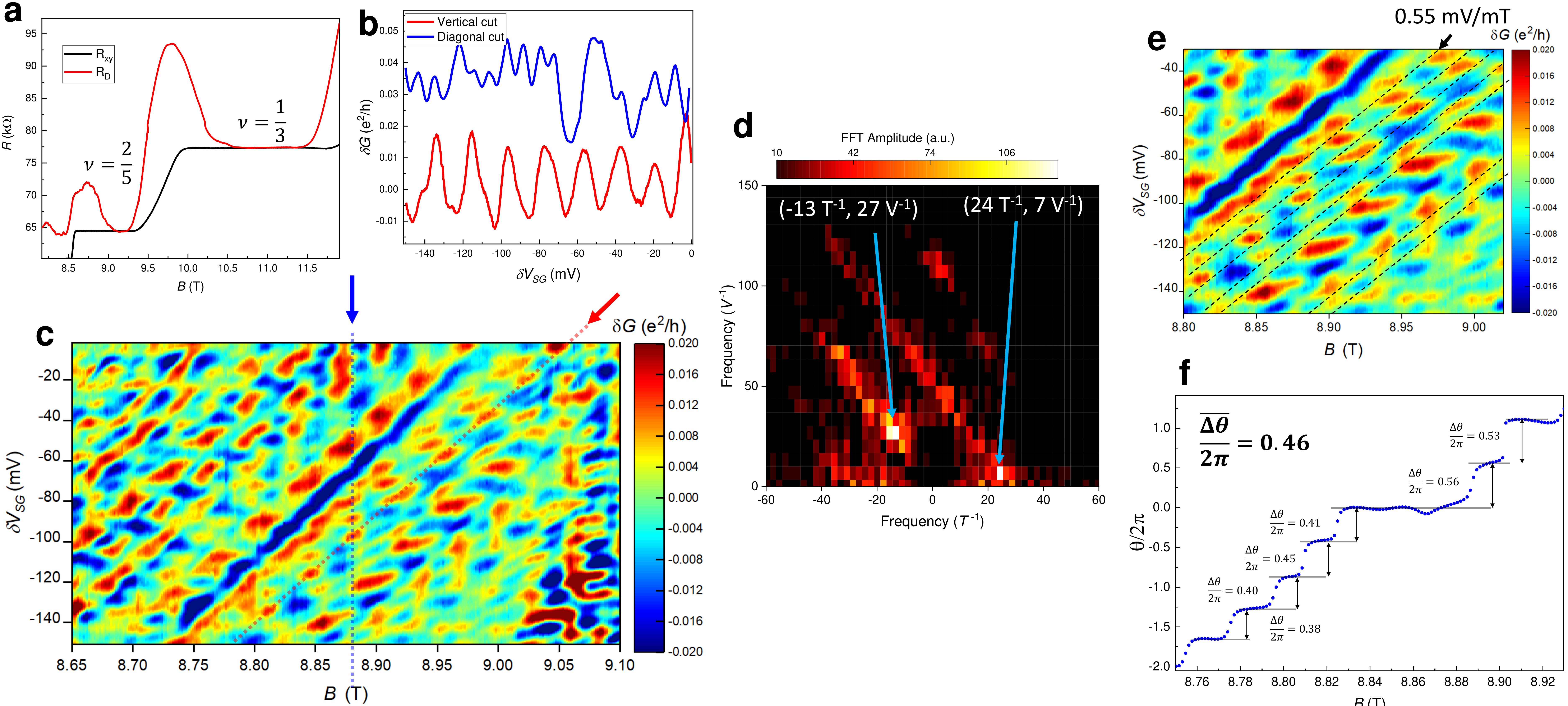}
\caption{\label{Device_B_Inner}Device B at $\nu = 2/5$. a) $R_D$ (red) and $R_{xy}$ measured with the 2DEG under the gates just depleted, but minimal backscattering, $V_{gates} = -0.3$ V. There is a wide region where $\nu = 1/3$ is fully transmitted and a much narrower region where $\nu = 2/5$ reaches close to full transmission of both edge states. b) Line cuts of interference of the inner mode at $\nu = 2/5$. The blue line is a vertical line cut, while the red line is a diagonal line cut. c)Interferece of the inner mode at $\nu = 2/5$. In addition to the interference oscillations, there are positively-sloped oscillations attributed to resonant tunneling through the bulk. Positions of the line cuts in b) are indicated with red and blue arrows and dashed lines. d) Two-dimensional FFT of the data from c. The peak at (-13 T$^{-1}$, 27 V$^{-1}$) corresponds to the positively-sloped oscillations, which we attribute to resonant tunneling, while the peak at (24 T$^{-1}$, 7 V$^{-1}$) corresponds to the interference process. (d) Zoom-in of the region with discrete jumps in the interference phase. Dashes lines indicate the positions of discrete jumps. e) Phase extracted via FFT versus magnetic field. Discrete jumps form a staircase pattern, with the values of the discrete jumps indicated. }
\end{figure*}

Device B has dimensions 800 nm $\times$ 800 nm, making it smaller than Device A (which is the primary device described in the main text). The Aharonov-Bohm period at $\nu = 1$ is 26 mT, indicating an effective interference area of $\approx 0.16$ $\mu$m$^2$, consistent with a $\approx200$ nm depleted region at the edge of the gates. 

Similar to Device A, with the gates just depleted Device B has full transmission at $\nu = 1/3$ in a wide region and full transmission at $\nu = 2/5$ in a narrow region, which can be seen from plots of $R_D$ and $R_{xy}$ in Supp. Fig. \ref{Device_B_Inner}a. With the device tuned to interfere the inner edge mode at 2/5, there is a complicated pattern, shown in Supp. Fig. \ref{Device_B_Inner}. It can be seen that some of the behavior has a clear positive slope and large magnetic field period, similar to the compressible-regime behavior of Device A. However, in line cuts of conductance along diagonal cuts in the $B-V_{SG}$ plane, there are nearly sinusoidal oscillations, shown in Supp. Fig. \ref{Device_B_Inner}b (red curve). Since this line cut is parallel to the positively-sloped oscillations that occur, the $\delta G$ along this contour does not oscillate with the positively-sloped contribution to the conductance variation. The fact that these sinusoidal oscillations still occur along the positively-sloped stripes suggests that, as discussed for Device A in the main text, the positively-sloped oscillations are caused by resonant tunnelling through the bulk, while the other oscillations are due to an interference process. It is noteworthy that both sets of oscillations are significantly larger in Device B than Device A, likely due to the approximately 2x smaller effective area. A smaller area is likely to enhance conduction through the bulk since the separation of the edges is smaller, while also increasing interference amplitude.

The co-existence of both types of oscillations creates a complicated pattern, particularly in the lower field region around 8.75 T where both are strong. In Supp. Fig. \ref{Device_B_Inner}d we show a two-dimensional Fourier transform. A prominent peak occurs at (-13 T$^{-1}$, 27 V$^{-1}$), which corresponds to the positively sloped oscillations that we attribute to resonant tunneling through the bulk. It is noteworthy that there are also contributions to the spectrum at higher frequencies with the same ratio of magnetic field to gate voltage frequency. The interference oscillations, on the other hand, create a nearly vertical pattern of oscillations, qualitatively similar to the compressible regime behavior in Supp. Fig. \ref{Simulations}a. This leads to the peak at (24 T$^{-1}$, 7 V$^{-1}$), with the small gate voltage frequency indicative of a very steep slope. The coexistence of bulk resonant tunneling and interference suggests that disentangling these effects may be challenging, particularly for states with small gaps and for small devices, and it is important to have conductance data as a function of both magnetic field and gate voltage.

Similar to Device A, there are discrete jumps in the phase of the interference oscillations, indicated with dashed lines in Supp. Fig. \ref{Device_B_Inner}d. The phase is plotted in Supp. Fig. \ref{Device_B_Inner}e, extracted via Fourier transform. Since the discrete jumps in phase occur across lines parallel to the positively-sloped resonant tunneling oscillations, the resonant tunneling oscillations do not significantly affect the extracted phases. As with Device A, in the regions between the discrete jumps the phase does not vary significantly as a function of magnetic field, suggesting the effects of bulk-edge coupling; therefore, we have not subtracted off a continuously varying AB slope. Compared to Device A, the discrete jumps for Device B have smaller spacing in terms of $B$, despite the fact that the device is smaller. There is one nearly-flat region of of $\approx 60$ mT extent which may correspond to a narrow incompressible bulk region, but above and below this narrow region there are jumps discrete jumps with spacing $\approx 17$ mT, corresponding to $\approx 0.65 \Phi_0$. This is close to the value of $\frac{\Phi_0}{2}$ that would be expected for the creation of quasiparticles with a fully compressible bulk; the fact that the spacing of the discrete jumps is slightly larger than this may suggest that the area of the inner puddle is slightly smaller than the effective area at $\nu = 1$, or that the bulk is not quite fully compressible. Nevertheless, this small period points to Device B exhibiting interference in a regime in which the bulk is nearly fully compressible. The fact that interference is measurable in this regime in Device B is consistent with the fact that the smaller area should increase the bulk charging energy, leading to less thermal smearing of the $e/5$ quasiparticle number.

The average value of the discrete jumps is $\Bar{\Delta \theta} = 0.46 \times 2 \pi$, close to (but slightly smaller than) the value for Device A.

The fact that Device B lacks a wide incompressible region may suggest that it is more disordered than Device A. Although both devices are fabricated on the same wafer and fabricated at the same time, the microscopic disorder potential will vary from device to device. With the energy gap at $\nu = 2/5$ being relatively small, it is plausible that it could be overwhelmed by disorder in some devices, giving minimal incompressible regime behavior at $\nu = 2/5$.

\begin{figure}[t]
\def\ffile{Outer_800nm_flat.pdf}
\centering
\includegraphics[width=1.0\linewidth]{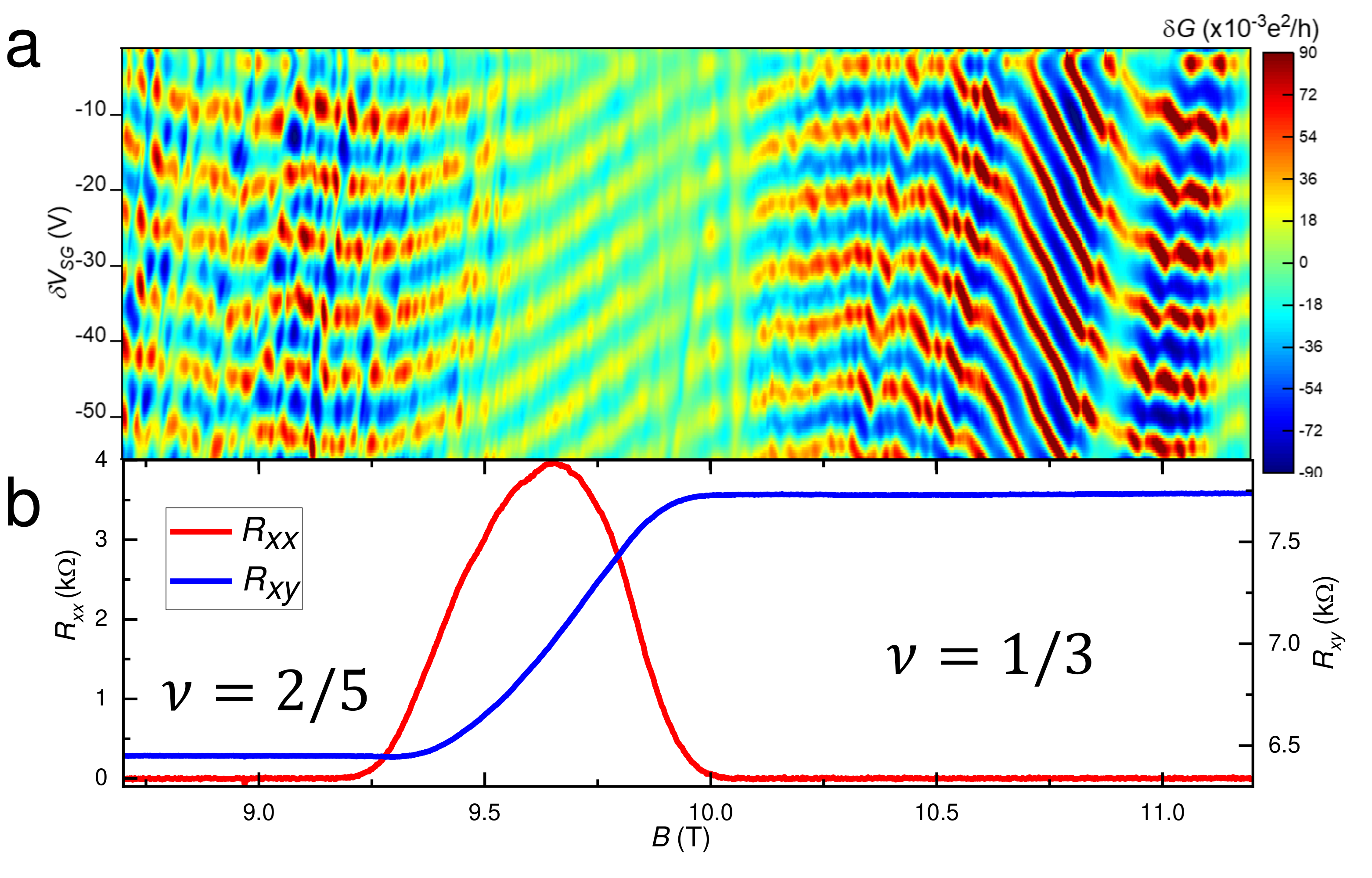}
\caption{\label{Device_B_Outer}Interference of device B outer mode from $\nu = 2/5$ to $\nu = 1/3$. a) Conductance oscillations versus $B$ and $\Delta V_{SG}$. Similar to device A shown in the main text, the interference is continuous from $\nu = 1/3$ to $\nu = 2/5$. Unlike device A, there is no clear incompressible region at $\nu = 2/5$, with the device seeming to exhibit compressible-regime interference across 2/5. b) Bulk magnetotransport $R_{xx}$ and $R_{xy}$.}
\end{figure}

Supp. Fig. \ref{Device_B_Outer} shows interference of the outer mode at $\nu = 2/5$ continuously measured up to $\nu = 1/3$. Similar to Device A, the oscillation pattern is continuous up to $\nu = 1/3$, though with suppressed amplitude in the intervening region where the bulk conductance is high. At $\nu = 1/3$, there is a negatively-sloped region with a few discrete jumps, consistent with previous observations of anyonic interference \cite{Nakamura2020, Nakamura2022}. The negatively-sloped incompressible region at $\nu = 1/3$ has a magnetic field extent of $\approx 400$ mT, somewhat narrower than Device A, supporting the possibility that Device B is more disordered. This is further reinforced by the fact that there is no clear negatively-sloped incompressible region at $\nu = 2/5$ for Device B. Since this region is already relatively narrow in Device A, it is plausible that the higher disorder level shrinks the incompressible region down to virtually zero extent at $\nu = 2/5$, which impacts interference of both the inner and outer modes.

\begin{figure}[t]
\def\ffile{nu_1_3_800nm_v2.pdf}
\centering
\includegraphics[width=1.0\linewidth]{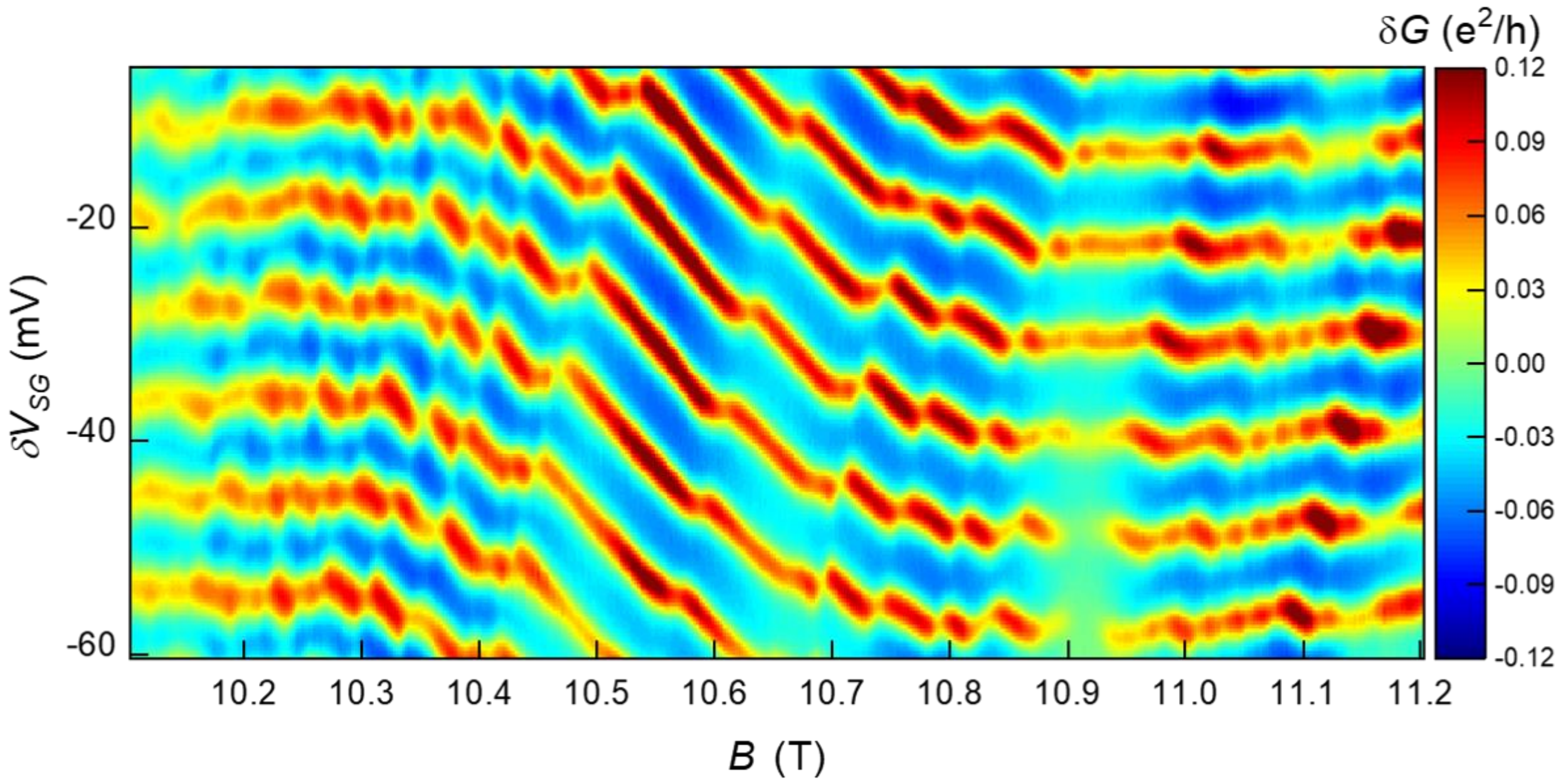}
\caption{\label{Device_B_1_3}Interference at $\nu = 1/3$ in device B. Conductance oscillations versus $B$ and $\delta V_{SG}$ at $\nu = 1/3$. Discrete jumps in phase are evident in the incompressible region.}
\end{figure}

Supp. Fig. \ref{Device_B_1_3} shows interference at $\nu = 1/3$ for Device B. Similar to Device A (and consistent with previous observations), there is an incompressible central region with a few discrete jumps in phase, with compressible regions at higher and lower field where the lines of constant phase are nearly flat. In the high and low field regions, there are periodic modulations with period $\approx \Phi_0$ due to periodic creation/removal of quasiholes/quasiparticles. These modulations are clearly visible due to the smaller size and larger charging energy of Device B; only weak hints of this behavior are visible for Device A in Supp. Fig. \ref{nu_1_3}. 

\begin{figure*}[t]
\def\ffile{DeviceC_data_flat_v2.pdf}
\centering
\includegraphics[width=1.0\linewidth]{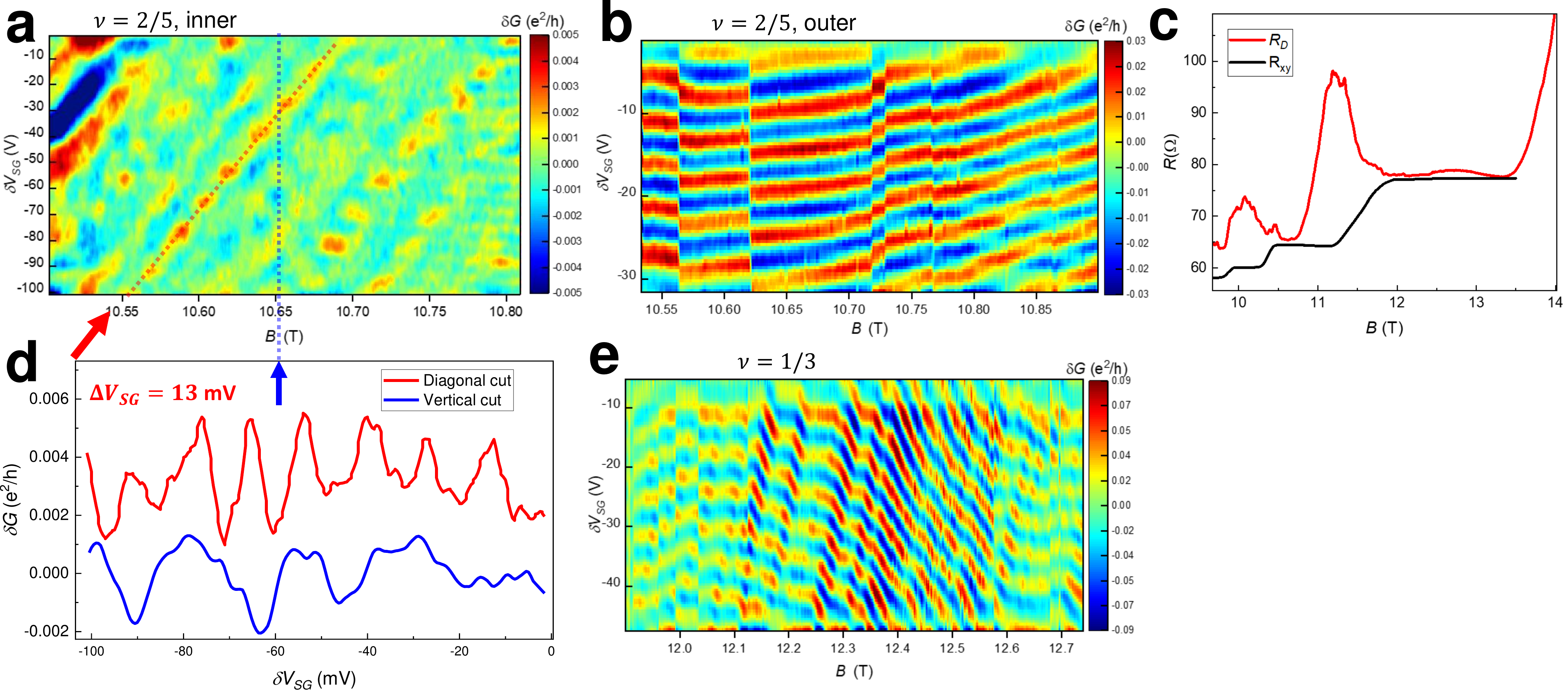}
\caption{\label{Device_C}Data from Device C. a) Conductance oscillations versus $B$ and $\delta V_{SG}$ for the inner mode at $\nu = 2/5$. b) Oscillations from the outer mode at 2/5. Several charge switching events are visible, with this device being less stable than devices A and B. c) $R_D$ and $R_{xy}$. Unlike Devices A and B, this device does not reach the quantized value of conductance at 2/5, as can be seen from the fact that the minimum resistance is still slightly higher than the $R_{xy}$ plateau. There is also a small amount of excess resistance at 1/3 in $R_D$. These features suggest that the device is more disordered. d) Vertical (blue) and diagonal (red) line cuts from a). Along the diagonal line cut, there are nearly sinusoidal oscillations with a gate voltage period of 13 mV, close to the value from Device A for the inner mode, but the amplitude is much smaller. e) Intereference at $\nu = 1/3$. The device shows qualitatively the same behavior as Devices A and B, but with additional noise both in the form of charge switching events and higher frequency noise, which is more visible at higher fields. Discrete jumps in phase are visible.}
\end{figure*}

\section{Supplementary Section 11: Device C}

Device C has lithographic dimensions $1$ $\mu$m $\times$ 1$\mu$m, the same as Device A, and is fabricated on the same wafer as Devices A and B. The only difference from Device A is that Device C is on a chip was taken from a region of the wafer farther from the center. The Aharonov-Bohm period at $\nu = 1$ is 11.4 mT, implying an effective area $\approx 0.36$ $\mu$m$^2$, close to the value for Device A.

Supp. Fig. \ref{Device_C}a shows conductance versus $B$ and $\Delta V_{SG}$ for the inner mode at $\nu = 2/5$. Despite having the same dimensions and heterostructure, in this device only weak hints of the interference behavior observed in Device A are visible. Along some diagonal contours, weak oscillations with nearly the same gate voltage period as device A occur; an example line cut is shown in Fig. \ref{Device_C}d (red). However, the amplitude is much smaller, making these oscillation barely resolvable in our measurements. Similar to the compressible regime behavior in Device A and Device B, there are some weak positively sloped conductance peaks, likely caused by resonant tunneling through the bulk.

Fig. \ref{Device_C}b shows oscillations for the outer mode at $\nu = 2/5$. The behavior is similar to the compressible-regime behavior for Device A and B for the outer mode. Unlike Device A, there is no indication of an incompressible region, suggesting that this device is more disordered. Additionally, there are several charge switching events; with the more negative QPC voltages required to backscatter the outer mode, we found that Device C becomes quite unstable.

Fig. \ref{Device_C} shows diagonal resistance $R_{D}$ and bulk Hall resistance $R_{xy}$ for Device C with the QPCs just depleted, $V_{gates} = -0.3$ V. Notably, at $\nu = 2/5$, $R_D$ shows a minimum where it approaches $R_{xy}$, but even at this minimum $R_D$ remains somewhat higher than $R_{xy}$, indicating that the edge states never fully transmit through the device. This is another indication that Device C is more disorded than Device A and B. At $\nu = 1/3$, $R_D$ comes close to the quantized value of $R = 3 \frac{h}{e^2}$, but still exhibits some excess resistance. 

Fig. \ref{Device_C}e shows interference at $\nu = 1/3$. Overall the device shows similar behavior to Devices A and B, consistent with previous experiments \cite{Nakamura2020, Nakamura2022}, including several discrete jumps in phase. There is also some excess noise, both in the form of charge switching events due to the lower stability of the gates in this device, as well as some higher frequency noise that becomes more substantial at higher fields. We do not know the origin of this high frequency noise, but we speculate that it might be related to the bulk conduction through the device implied by the excess resistance in $R_{D}$ at $\nu = 1/3$ visible in Fig. \ref{Device_C}. It also noticeable that the incompressible region is significantly narrower than in Device A, approximately 350 mT compared to 600 mT in Device A.

The data at $\nu = 2/5$ indicates that interference is degraded in Device C compared to Device A, most likely due to a higher level of disorder. The higher disorder level could be because the device is closer to the edge of the wafer, or could simply be a random effect due to small inconsistencies in nanofabrication and the specific disorder potential created by ionized Si donors. Interference of the inner mode at $\nu = 2/5$ is almost completely destroyed, while interference at $\nu = 1/3$ is still present and reasonably clear, but with more noise and a somewhat narrower incompressible gapped region. This gives experimental support to the theoretical expectation from Ref. \cite{Feldman2022} that interference should be especially robust at $\nu = 1/3$ (as well as the integer state $\nu = 1$) where there is a single chiral edge state. The large intrinsic gap of $\nu = 1/3$ should also contribute to this robustness.

\bibliography{Interferometer_Bib}